\documentclass[12pt,a4paper]{article}
\usepackage[T1]{fontenc}
\usepackage[utf8]{inputenc}
\usepackage{authblk}
\usepackage{bbding} 

\usepackage{amsmath}
\usepackage{graphicx,float,color}
\usepackage{hyperref}
\hypersetup{colorlinks = true, linkcolor = blue, urlcolor  = blue, citecolor = blue, anchorcolor = blue}
\usepackage{multirow}
\usepackage[numbers,sort&compress]{natbib}
\usepackage{braket}
\usepackage{amsfonts}
\usepackage{amssymb} 
\usepackage{epic}
\usepackage{eepic}
\usepackage{epsfig}
\usepackage{latexsym}
\usepackage{color}
\usepackage{url}  
\usepackage{caption}
\usepackage[caption=false]{subfig}
\usepackage{float}
\usepackage{bm}
\usepackage{geometry}
\usepackage{slashed}
\usepackage[detect-all]{siunitx}

\def\lp{\ell_{\mathrm{Pl}}}

\usepackage{lipsum}
\makeatletter
\def\ps@pprintTitle{%
 \let\@oddhead\@empty
 \let\@evenhead\@empty
 \def\@oddfoot{}%
 \let\@evenfoot\@oddfoot}
\makeatother
\usepackage{etoolbox}
\begin{document}
\title{\textbf{Gravitational wave echoes search with combs}}    

\author[1,2]{Jing Ren\thanks{renjing@ihep.ac.cn}}
\author[1,2]{Di Wu\thanks{wudi@ihep.ac.cn}}
\affil[1]{\normalsize Institute of High Energy Physics, Chinese Academy of Sciences, Beijing 100049, China}
\affil[2]{\normalsize School of Physics Sciences, University of Chinese Academy of Sciences, Beijing 100039, China}

\maketitle

%\date{\today}
%\vspace{0.6cm}
\begin{abstract}
Gravitational wave echoes may provide a smoking gun signal for new physics in the immediate vicinity of black holes. As a quasiperiodic signal in time, echoes are characterized by the nearly constant time delay, and its precise measurement can help reveal a Planck-scale deviation outside of the would-be horizon. Different search methods have been developed for this quasiperiodic signal, while the searches suffer from large theoretical uncertainties of the echo waveform associated with the near-horizon physics. On the other hand, a coherent combination of a large number of pulses gives rise to a generic narrow resonance structure for the echo amplitude in frequency. This quasiperiodic pattern sets a complementary search target for echoes, and the time delay is inversely related to the average resonance spacing. A uniform comb has been proposed to look for the resonance structure in a rather model-independent way. In this paper, we develop a Bayesian algorithm to search for the resonance structure based on combs, where a phase-marginalized likelihood plays an essential role. The algorithm is validated with signal injections in detector noise from Advanced LIGO. With special treatments of the non-Gaussian artifacts, the noise outliers of the log Bayes factor distribution are properly removed. An echo signal not significantly below noise is detectable, and the time delay can be determined to very high precision. We perform the proposed search on real gravitational wave strain data  of the first observing run of Advanced LIGO. We find no clear evidence of a comblike structure for GW150914 and GW151012. \\
\\
\textit{Keywords: gravitational wave echoes, time delay, resonance structure, Bayesian algorithm, combs, phase-marginalized likelihood, instrumental lines} 
\end{abstract}

%\maketitle
%%%%%%%%%%%%%%%%%%%%%%%%%%%%%%%%%
%To change the margin just for this page put the \newgeometry thing below right above the title and put the \restoregeometry in the end of the page-you gotta adjust that last part by hand. This is for that identified part arXiv put in the left hand side
%%%%%%%%%%%%%%%%%%%%%%%%%%%%%%%%%%%%%%%%%%%%%%%
%\newgeometry{left=1.8cm,right=1.5cm,bottom=2.5cm,top=2.8cm} 
\newpage
{
  \hypersetup{linkcolor=black}
  \tableofcontents
}
%\tableofcontents
%\vspace{0.5cm}
%\hrulefill
%\vspace{1cm}

\section{Introduction}

The LIGO, Virgo, and KAGRA Collaborations have detected gravitational wave signals from compact binary coalescences for more than 50 events~\cite{GWTC1, GWTC2}. The observations are consistent with general relativity (GR) predictions for black holes, and deviations at a distance of the order of the horizon size have been strongly constrained~\cite{LIGOScientific:2019fpa, Abbott:2020jks}. However, these observations tell little about physics in the immediate vicinity of black holes, where strong deviations may make their first appearance. Near-horizon corrections might be fundamental to resolving the black hole information paradox~\cite{tHooft:2016fzb}. Ultracompact objects (UCOs), in particular, are an interesting alternative, where many theoretical candidates have been proposed~\cite{Cardoso:2019rvt}. As one extreme possibility, the horizon as a one-way membrane could be fully removed, and gravitational waves falling into UCOs may have chance to be reflected out. This is not in contradiction with the electromagnetic observations of  astrophysical black holes as dark objects. With the wavelength much shorter than the objects' size, the electromagnetic luminosity will be highly suppressed by the efficient trapping in the high redshift region. 

\begin{figure}[!h]
  \centering%
{ \includegraphics[width=15cm]{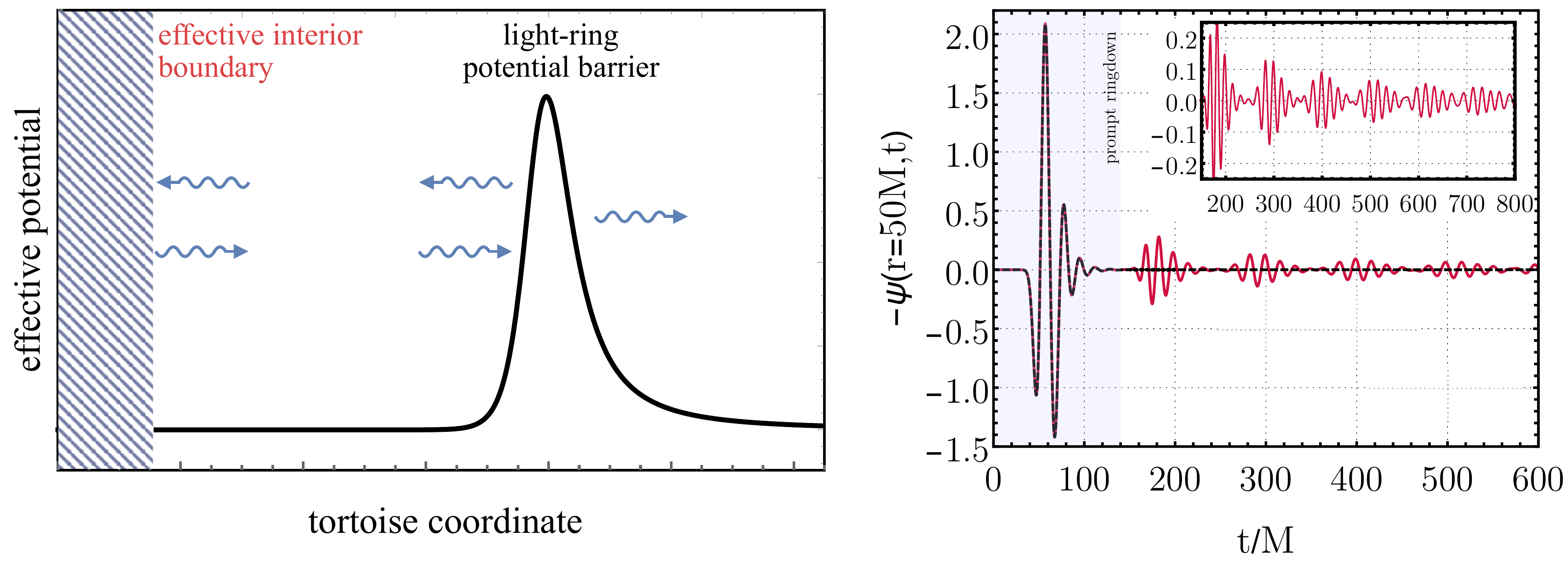}}
\caption{\label{fig:echo1} Left: a schematic plot of the leaky cavity for the propagation of gravitational waves. Right (adapted from Ref.~\cite{Cardoso:2019rvt}): the postmerger waveform for a black hole (dashed black curve) and for a UCO with echoes (solid red curve). }
\end{figure}

If a substantial reflection is allowed, gravitational waves can be considered as propagating inside a leaky cavity bounded by the light-ring potential barrier and an effective interior boundary, as demonstrated in the left panel in Fig.~\ref{fig:echo1}. The former has been well probed by the current observations of a clear black hole ringdown signal~\cite{Cardoso:2016rao}, and the latter encodes the essential information for near-horizon corrections. 
Echoes then originate from a repeated and damped reflection of the initial pulse inside the cavity that gradually leaks out through the light-ring potential barrier~\cite{Cardoso:2016oxy}. This produces a quasiperiodic signal in time after the dominant black hole ringdown, as shown in the right panel in Fig.~\ref{fig:echo1}. Echoes are characterized by a nearly constant time delay $t_d$ between two consecutive pulses.
Considering a simple model of UCOs, a Kerr black hole truncated at a radius $r_0$ slightly outside the would-be horizon, the time decay is~\cite{Cardoso:2019rvt}
\begin{eqnarray}\label{eq:timedelay}
\frac{t_d}{M}\approx -2\,\ln\epsilon \left[1+(1-\chi^2)^{-1/2}\right]
\equiv 2\,\eta \left(\ln\frac{M}{\lp}\right) \left[1+(1-\chi^2)^{-1/2}\right]\,,
\end{eqnarray}
where $M$ and $\chi$ are the mass and dimensionless spin of the object, respectively, and $\epsilon=r_0/(2M)-1$. $\ln\epsilon$ denotes the generic logarithmic dependence on the small distance from the would-be horizon to where deviations occur. It is convenient to use $\eta\equiv -\ln\epsilon/\ln(M/\lp)$ to quantify the smallness of the distance. $\eta=1$ is for a coordinate Planck distance ($r_0-2M\sim\lp$), and $\eta=2$ means that the proper distance for $r_0-2M$ is of the Planck length. 
For $M\sim 10 M_\odot$, $t_d$ is of the order of 0.1\,s for $\eta\sim\mathcal{O}(1)$ and is quite accessible  with current experiments. 
Thus, gravitational wave echoes serve as a perfect target to search for strong deviations at a microscopic distance outside of macroscopic objects.

\begin{figure}[!h]
  \centering%
{ \includegraphics[width=14.5cm]{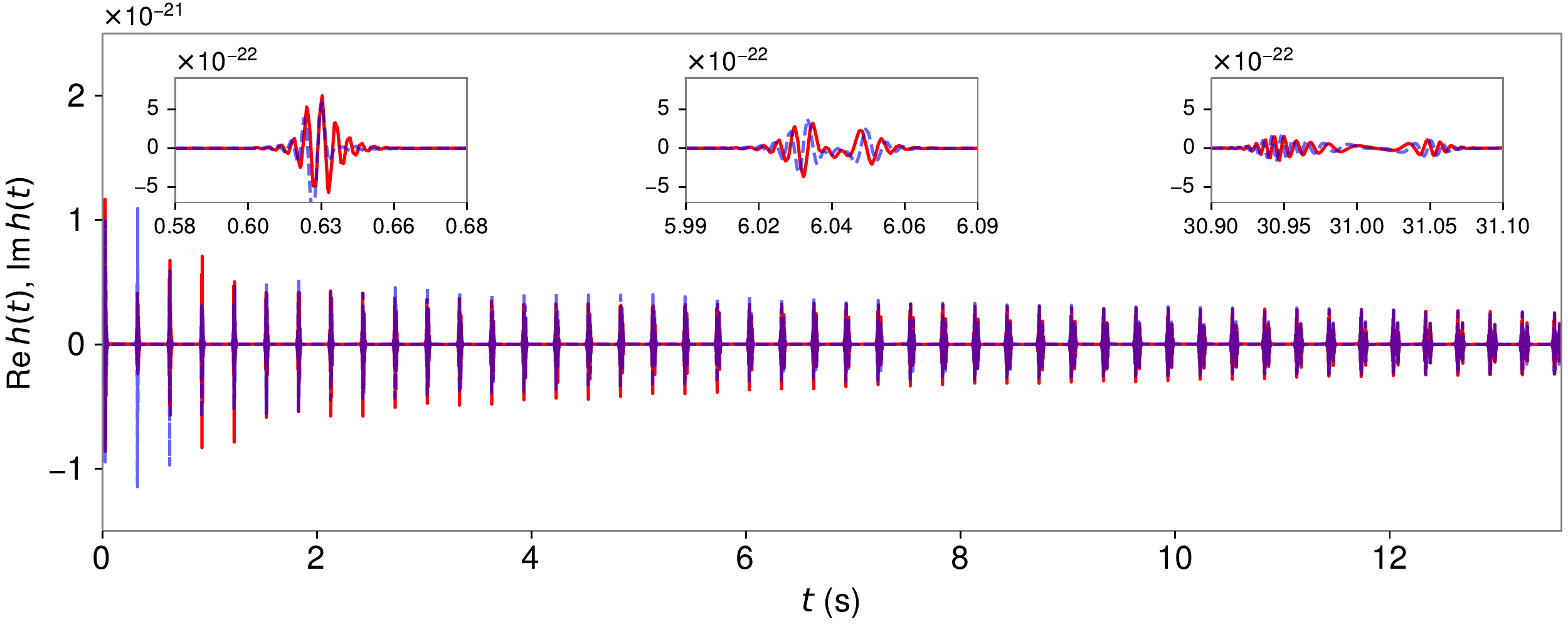}}\\
{ \includegraphics[width=14.3cm]{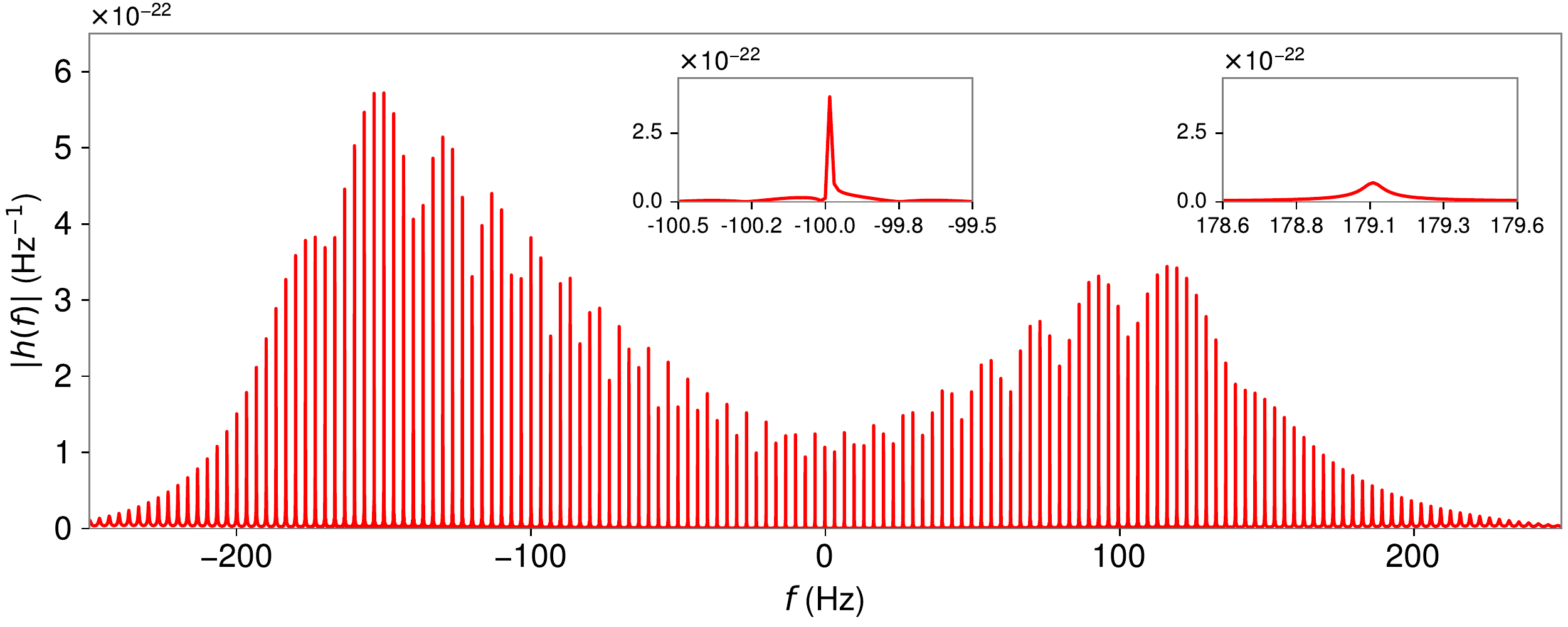}}
\caption{\label{fig:echo2} Top: the real (solid red line) and imaginary (dashed blue line) parts of an analytical model of the echo waveform for a truncated Kerr black hole with $M=69\,M_\odot$, $\chi=0.7$, $t_d= 0.3\,$s~\cite{Maggio:2019zyv}. The insets show the profiles of the 3rd, $\sim 20$th and $\sim100$th pulse from left to right, respectively. Bottom: the corresponding echo amplitude in the frequency domain by incorporating about 200 pulses. It features a narrow resonance structure asymmetric at positive and negative frequencies. The insets show the examples of narrower and wider resonances.}
\end{figure}

Template-based search methods have been actively developed to search for the quasiperiodic  signal of echoes in the time domain~\cite{Abedi:2016hgu, Nielsen:2018lkf, Lo:2018sep}. This requires a careful modeling of the echo waveform, in particular for spinning UCOs~\cite{Nakano:2017fvh, Wang:2018gin, Maggio:2019zyv, Xin:2021zir}. However, the waveform has large theoretical uncertainties due to largely unknown details associated with UCOs, e.g. exact shape of the cavity, nonperturbative physics, and reflection of the effective interior boundary \cite{Wang:2018mlp, Conklin:2019fcs, Oshita:2019sat, Chen:2020htz, Srivastava:2021uku}. The top panel in Fig.~\ref{fig:echo2} shows an example of the echo waveform for a truncated Kerr black hole model. The pulse profile varies significantly with time, and the phase evolution that plays the essential role in matched filtering suffers, in particular, from the theoretical uncertainties. So far, only a few toy models of the echo waveform have been thoroughly explored and applied to the real data search. No clear evidence of postmerger echoes has been found for confirmed events in  the GWTC-1 and GWTC-2 catalogs~\cite{Westerweck:2017hus, Uchikata:2019frs, Abbott:2020jks, Wang:2020ayy}, although some tentative evidences were  reported initially~\cite{Abedi:2016hgu, Abedi:2018npz, Abedi:2020sgg}. 

The null results could be related to a potentially large mismatch between the realistic waveform and the templates. 
Thus, it is necessary to develop less model-dependent search methods that target the essential features of echoes regardless of details. 
A morphology-independent search of a quasiperiodic  signal in time has been proposed based on the BayesWave algorithm~\cite{Tsang:2018uie, Tsang:2019zra}. It models each pulse separately by a sine-Gaussian and, thus, is quite generic. The method could be powerful if the signal power is dominated by a few early pulses with large amplitudes and clear shapes. However, if the initial condition is dominated by the low-frequency modes, the search has to include many pulses, and the method will quickly become inefficient with a dramatically increasing number of parameters.  A different search target has to be considered. 

It turns out that a coherent combination of a large number of pulses leads to a narrow resonance structure for the echo amplitude in the frequency domain, as shown by the bottom panel in Fig.~\ref{fig:echo2}. Ignoring the phase information, the resonance structure has quite generic features. When the cavity is sufficiently long, the narrow resonances are quasiperiodic , corresponding to the long-lived quasinormal modes of the cavity~\cite{Mark:2017dnq}. The average spacing between resonances is inversely related to $t_d$ and can be used to infer microscopic deviations around the would-be horizon. Narrow resonances appear with the absolute value of frequency $|f|$ below the dominant black hole ringdown frequency, and they become narrower at lower $|f|$ if there is no strong damping from the interior. For a given frequency resolution, the resonances at small $|f|$ are too narrow to be resolved, and the pattern declines again for $|f|$ below certain values. 
These generic features apply to resonances at both positive and negative frequencies that are generally asymmetric for a spinning object. 
Thus, the resonance structure provides a robust search target for echoes that suffers less from the theoretical uncertainties.  It is complementary to the aforementioned search methods that focus on early time pulses, providing a unique way to extract the time delay when the initial condition is dominated by low-frequency modes and the early pulses are  not well separated.

A uniform comb has been proposed to capture the dominant contribution of the resonance structure~\cite{Conklin:2017lwb}. As a crude model of echo amplitude, the comb consists of a large number of teeth with equal height, width, and spacing and can be defined by a small number of parameters. 
Based on this idea, simple search strategies with a uniform comb have been developed, and useful test statistics are constructed to search for the resonance structure~\cite{Holdom:2019bdv}. Applying the search to real data, tentative evidences for echoes have been reported for ten events in the GWTC-1 catalog, with interesting correlation of the inferred parameter $\eta$ in Eq.~(\ref{eq:timedelay}) for the microscopic derivation. 
However, the method is not fully automatized, and the results have not yet been reproduced by others. 
Further optimizations of the method are required. 

In this paper, we develop a Bayesian algorithm to search for the narrow resonance structure based on combs. 
To model echo amplitude in the frequency domain without phases, we employ a phase-marginalized likelihood and consider combs as templates for the generic resonance structure. The method is not expected to recover signals much below the average noise due to the lack of phase information, but its generic features enable detections of a large variety of echo waveforms in a rather model-independent way. The algorithm allows a coherent combination of multiple detectors, which has not been considered in the previous study with combs~\cite{Holdom:2019bdv}, and this will further enhance the detection probability. 
In Sec.~\ref{sec:method}, we describe our Bayesian algorithm, with the phase-marginalized likelihood and the comb model as the essential inputs. 
The parameter settings are discussed for both intrinsic and extrinsic parameters. 
Section~\ref{sec:result} displays the search results. We first validate the algorithm with signal injections in stationary Gaussian noise to understand the behavior of phase-marginalized likelihood. Then, we apply the proposed search to real gravitational wave strain data of the first observing run of Advanced LIGO. The narrow instrumental lines are the main non-Gaussian artifacts of concern for the narrow resonances search. 
We analyze the background distribution of the analysis, and identify a small number of prominent lines to be mitigated to improve the search sensitivity. 
The search results for GW150914 and GW151012 are shown in the end. We summarize in Sec.~\ref{sec:summary}.

\section{Method}
\label{sec:method}

We perform Bayesian model selection and parameter estimation for postmerger echoes on confirmed gravitational wave events.
A resonance structure not much below the average noise can be recovered by using the phase-marginalized  likelihood function and a comb model. 
We consider the log Bayes factor for the signal versus noise models as the detection statistic to decide whether there is a comblike signal or not. 
When a signal is identified, the inferred comb parameters provide valuable information for the resonance structure. 
In the following, we first derive the generic phase-marginalized likelihood function, and then discuss the parameter settings specific to the echo search.

\subsection{Phase-marginalized likelihood}
\label{sec:likelihood}

Take $d=n+h$ to represent the Fourier transform of a strain time series of duration $T$. The noise $n$ is typically assumed to be stationary and Gaussian, and $h$ denotes the gravitational wave signal. The  likelihood for the data in a single frequency bin $j$ given the signal model $h_j$ is then $\mathcal{L}(d_j | h_j)=\frac{1}{2\pi P_j}\exp\left(-2\, \delta f \,|d_j-h_j|^2/P_j\right)$, where $\delta f=1/T$ is the frequency resolution and $P_j$ is the known one-sided power spectrum density (PSD) for noise. 

The likelihood used for estimation of the signal amplitude can be derived by marginalizing over the signal model phase of the Gaussian likelihood. In the polar coordinate, the Gaussian likelihood becomes
\begin{eqnarray}\label{eq:GaussianL1}
\mathcal{L}(d_j | h_j)=\frac{|d_j|}{2\pi P_j}\exp\left(-\frac{1}{2}\frac{|d_j|^2+|h_j|^2}{\tilde{P}_j}\right)
\exp\left(\frac{|d_j||h_j|}{\tilde{P}_j} \cos(\phi_{j}-\delta_{j}) \right),
\end{eqnarray}
where $\tilde{P}_j=P_j/(4 \,\delta f)$ has the unit of Hz$^{-2}$ and $\phi _{j}=\arg(d_j)$ and $\delta _{j}=\arg(h_j)$ are phases of the data and signal model, respectively. The phase-marginalized likelihood is then~\cite{Ashton:2018jfp}
\begin{eqnarray}\label{eq:likelihoodsingle}
&&\mathcal{L}(d_j | |h_j|)=\int_0^{2\pi} \mathcal{L}(d_j | |h_j|, \delta_{j}) \,\pi(\delta_{j}) \, d\delta_{j}\nonumber\\
&=&\frac{|d_j|}{2\pi P_j}\exp\left(-\frac{1}{2}\frac{|d_j|^2+|h_j|^2}{\tilde{P}_j}\right)I_0\left(\frac{|d_j||h_j|}{\tilde{P}_j}\right),
\end{eqnarray} 
where a flat prior for the signal model phase $\delta_{j}$ is assumed, i.e., $\pi(\delta_j)=1/(2\pi)$, and 
$I_0$ is the zeroth-order modified Bessel function of the first kind.\footnote{The phase-marginalized likelihood is called the power-spectral density likelihood in Ref.~\cite{Ashton:2018jfp}. But the expression in Eq.~(5.2) has a typo, and $\ln|h_i|$ in the second line should be $\ln|d_i|$. We thank Paul Easter for confirming this in a private communication.} Note that Eq.~(\ref{eq:likelihoodsingle}) has no dependence on the data phase $\phi_j$ either, and it is just the Rice distribution of $|d_j|$ up to a constant. Combining different frequency bins, the marginalized log-likelihood for the data $d$ given the signal model amplitude $|h|$ parametrized by $\theta$ is
\begin{eqnarray}\label{eq:logL1}
\ln\mathcal{L}(d | \theta)=\ln\mathcal{Z}_0+\rho_\textrm{mf}\, \rho_\textrm{opt}-\frac{1}{2}\rho_\textrm{opt}^2\,,
\end{eqnarray}
where $\ln\mathcal{Z}_0=\sum_{j=1}^N \ln\mathcal{L}(d_j | 0)$ denotes log-evidence for noise and 
\begin{eqnarray}
\rho_\textrm{opt}^2=\sum_{j=1}^N \frac{|h_j|^2}{\tilde{P}_j},\;\;
\rho_\textrm{mf}=\frac{1}{\rho_\textrm{opt}}\sum_{j=1}^N \ln I_0\left(\frac{|d_j||h_j|}{\tilde{P}_j}\right).
\end{eqnarray}
$\rho_\textrm{opt}$ is the conventional optimal signal-to-noise ratio (SNR). $\rho_\textrm{mf}$ quantifies the overlap between the signal and data with no phase information.
The main effects from marginalizing over phases can be seen by approximating the log-Bessel function at the leading order: 
\begin{eqnarray}\label{eq:logI0}
\ln I_0(x)=\left\{
\begin{array}{ll}
x-\frac{1}{2}\ln(2\pi x),& x\geq 2\\
\frac{1}{4}x^2, &  x<2\,.
\end{array}\right.
\end{eqnarray}
When the  signal amplitude is larger than the average noise, $\rho_\textrm{mf}$ reduces mostly to the matched filter SNR that is linear in $|h|$, and the maximum likelihood is obtained when $|h|\approx |d|$. When the signal amplitude is small, $\rho_\textrm{mf}\,\rho_\textrm{opt}$  and $\rho_\textrm{opt}^2$ are both proportional to $|h|^2$, and the best-fit $|h|$ is driven to the average noise level. Thus, with no phase information, the method will largely lose sensitivity to small signals well below the average noise.

The search sensitivity can be significantly improved by a coherent combination of multiply detectors as in the case of stochastic gravitational wave backgrounds (SGWB) search. In this study, we consider a network of two detectors for the two LIGO interferometers with the best sensitivities. Generalization to more detectors is straightforward. 
Because of the correlated signal in different detectors, we shall first combine the likelihoods of two detectors at the same frequency bin, and then marginalize the combined likelihood over the common signal phase.

Ignoring the difference in detector responses first, the marginalized log-likelihood $\ln\mathcal{L}(\mathbf{d} | \theta)$ for two datasets $\mathbf{d}=\{d_I, d_J\}$ takes the same form as in Eq.~(\ref{eq:logL1}), with $\ln\mathcal{Z}_0=\ln\mathcal{Z}_{0,I}+\ln\mathcal{Z}_{0,J}$ and
\begin{eqnarray}\label{eq:rhomf2}
\rho^2_{\textrm{opt}}=\rho_{\textrm{opt},I}^2+\rho_{\textrm{opt},J}^2,\quad
\rho_{\textrm{mf}}=\frac{1}{\rho_{\textrm{opt}}}\sum_{j=1}^N \ln I_0\left(\, |h_j| \left|\frac{d_{I,j}}{\tilde{P}_{I,j}}+\frac{d_{J,j}}{\tilde{P}_{J,j}}\right|\right)\,.
\end{eqnarray}
As expected, a coherent combination preserves the relative phase between two datasets, and the cross-detector signal correlations are properly taken into account. The implication is more clear if we rewrite the SNRs in a more compact form. Defining a combined PSD $P_{IJ}=(P_I^{-1}+P_J^{-1})^{-1}$, Eq.~(\ref{eq:rhomf2}) can be written as
\begin{eqnarray}\label{eq:rhomf2p}
\rho^2_\textrm{opt}=\sum_{j=1}^N |\tilde{h}_j|^2,\;\;
\rho_\textrm{mf}=\frac{1}{\rho_\textrm{opt}}\sum_{j=1}^N \ln I_0\left(|\tilde{d}_{IJ,j}| |\tilde{h}_j|\right),
\end{eqnarray}
where $\tilde{h}=h/\sqrt{\tilde{P}_{IJ}}$ is the normalized signal and $\tilde{d}_{IJ}=\tilde{d}_{I}\cos\theta+\tilde{d}_{J}\sin\theta$ is the normalized combined data, with $\tilde{d}_{I,J}=d_{I,J}/\sqrt{\tilde{P}_{I,J}}$ and $\cos\theta = \sqrt{P_{IJ}/P_I}$. The likelihood for two detectors then takes the same form as that for one detector except that the PSD and the dataset are replaced by the combined versions. The sensitivity is then improved with $P_{IJ}$ smaller than that for a single detector. When $P_{I}$ and $P_{J}$ are comparable, the signal reach for the echo amplitude can be $\sqrt{2}$ times smaller. 
In contrast, an incoherent combination of two detectors leads to $\rho_\textrm{mf}\propto \ln I_0(|h||d_I|/\tilde{P}_I) + \ln I_0(|h||d_J|/\tilde{P}_J)$ with no cross-correlation. The search sensitivity is then worse. The difference between coherent and incoherent combinations will be more significant when more detectors are available.

Next, we incorporate the detector response effects. Gravitational wave detectors respond differently to signals depending on geometries of detectors and signal properties.  The detector response for a pointlike source can be derived from the impulse response $R^{ab}_I(f,\hat{n})$, where $f$ is the frequency and $\hat{n}$ denotes the unit vector pointing in the direction of the source.   
For Earth-based interferometers operating in the small-antenna limit, the response can be well approximated as~\cite{Romano:2016dpx}
\begin{eqnarray}\label{eq:RIab}
R^{ab}_I(f,\hat{n}) \approx \frac{1}{2}\left(u^a_Iu^b_I-v^a_I v^b_I\right) e^{i \,2\pi f\, \hat{n}\cdot \vec{x}_I/c}.
\end{eqnarray}
The magnitude is determined by the unit vectors $u_I$ and $v_I$ along the two arms of the detector $I$. The phase depends on the frequency, the source direction, and the detector vertex $\vec{x}_I$. Summing up two different polarizations, the detector response 
\begin{eqnarray}\label{eq:response}
h_I(f)=\frac{1}{2}\left[R_I(f) h(f)+R^*_I(-f) h^*(-f)\right]\,,
\end{eqnarray}
with $R_I(f)$ the effective impulse response. For a spinning object, the resonance structure is asymmetric at positive and negative frequencies as in Fig.~\ref{fig:echo2}. The projection in Eq.~(\ref{eq:response}) then ensures a real response in the time domain, and all information is encoded in $h_I(f)$ at positive frequency, where the resonances have a two-component structure~\cite{Conklin:2019fcs}. The search with a uniform comb picks out the dominant component with a larger amplitude. 
Marginalizing over the signal model phase,   the optimal SNR $\rho^2_{\textrm{opt},I}=\sum_{j=1}^N |R_{I,j}|^2|h_{j}|^2/\tilde{P}_{I,j}$, and the overlapping term in the combined likelihood function becomes 
\begin{eqnarray}\label{eq:rhomf3}
\rho_{\textrm{mf}}=\frac{1}{\rho_{\textrm{opt}}}\sum_{j=1}^N \ln I_0\left(|h_j|\, \left|\frac{d_{I,j}R_{I,j}^*}{\tilde{P}_{I,j}}+\frac{d_{J,j}R_{J,j}^*}{\tilde{P}_{J,j}}\right|\right).
\end{eqnarray}
The additional phases induced by detector response are properly removed by $R_{I,J}^*$, and this enables a coherent combination of the signals in two detectors. 

The precise form of $R_I$ is largely undetermined from the current analysis of the inspiral-merger-ringdown signal due to strong degeneracies of the extrinsic parameters~\cite{GWTC1, GWTC2}. Fortunately, the impact on our analysis is limited.  
Given the simple form of $R_I$ in Eq.~(\ref{eq:RIab}), the network optimal SNR and the overlapping term for two LIGO detectors are, respectively, 
\begin{eqnarray}\label{eq:rhomf4}
\rho^2_{\textrm{opt}}&=&\sum_j (|h'_j|^2/\tilde{P}_{H,j}+ A_{HL}^2 |h'_j|^2/\tilde{P}_{L,j})\nonumber\\
\rho_{\textrm{mf}} &=&\frac{1}{\rho_{\textrm{opt}}}\sum_{j=1}^N \ln I_0\left(|h'_j|\, \left|\frac{d_{H,j}}{\tilde{P}_{H,j}}+A_{HL} e^{i \phi_{HL,j}}\frac{d_{L,j}}{\tilde{P}_{L,j}}\right|\right),
\end{eqnarray}  
where $h'$ denotes the effective signal amplitude for one detector and the constant $A_{HL}$ parametrizes the amplitude ratio. The overlapping term depends on a relative phase in addition:
\begin{eqnarray}\label{eq:phiHL}
\phi_{HL,j}=\phi_{HL,0}-2\pi f_j \Delta t_{HL}\,,
\end{eqnarray}
where $\phi_{HL,0}$ denotes the constant term and the frequency-dependent part relies on the lag of the signal arrival time $\Delta t_{HL}=t_{H}-t_L$.  Since the time lag $\Delta t_{HL}$ is well determined from the current analysis and the constant $A_{HL}$, $\phi_{HL,0}$ distribute in narrow ranges due to the nearly antialigned configurations of the two LIGO detectors, the detector response-related uncertainties cause fewer problems here. 
Similarly, the combined likelihood with detector response can be put in a compact form, and Eq.~(\ref{eq:rhomf2p}) becomes
\begin{eqnarray}\label{eq:rhomf2res}
\rho^2_\textrm{opt}=\sum_{j=1}^N |\tilde{h}'_j|^2,\;\;
\rho_\textrm{mf}=\frac{1}{\rho_\textrm{opt}}\sum_{j=1}^N \ln I_0\left(|\tilde{d}_{HL,j}| |\tilde{h}'_j|\right),
\end{eqnarray}
where $P_{HL}=(P_H^{-1}+A_{HL}^2 P_L^{-1})^{-1}$ is the combined PSD, $\tilde{h}'=h'/\sqrt{\tilde{P}_{HL}}$ is the normalized signal and $\tilde{d}_{HJ}=(d_{H}/\tilde{P}_{H}+A_{HL} e^{i \phi_{HL}}d_{L}/\tilde{P}_{L})\sqrt{\tilde{P}_{HL}}$ is the normalized dataset.

To compute the log Bayes factor  for the signal versus noise models, it is convenient to define a normalized log-likelihood. The evidence for the signal model is 
\begin{eqnarray}
\mathcal{Z}_1=\int \mathcal{L}(\mathbf{d}|\theta)\, \pi(\theta) \,d\theta
=\mathcal{Z}_0 \int \exp\left(\rho_\textrm{mf}\, \rho_\textrm{opt}-\frac{1}{2}\rho_\textrm{opt}^2\right) \pi(\theta) d\theta\,,
\end{eqnarray}
where $\pi(\theta)$ is the priors for model parameters and $\mathcal{Z}_0$ is the noise evidence. The log Bayes factor is then
\begin{eqnarray}
\ln\mathcal{B}=\ln\mathcal{Z}_1-\ln\mathcal{Z}_0=\ln \int \tilde{\mathcal{L}}(\mathbf{d} | \theta)\, \pi(\theta)\, d\theta\,,
\end{eqnarray}
where the normalized log-likelihood is
\begin{eqnarray}\label{eq:logL2}
\ln\tilde{\mathcal{L}}(\mathbf{d} | \theta)=\rho_\textrm{mf}\, \rho_\textrm{opt}-\frac{1}{2}\rho_\textrm{opt}^2\,.
\end{eqnarray}
It has very clear meaning, reflecting the competition between the overlapping term and the network optimal SNR.

\subsection{Parameter settings}
\label{sec:3}

Bayesian inference automatically avoids overfitting with more complicated models. To optimize the search sensitivity, we need to choose a proper set of search parameters that maximizes the data and model overlap with the smallest possible number of parameters. There are two sets of parameters for our echo search: One specifies the comb that models the echo amplitude $|h|$ in Sec.~\ref{sec:likelihood},  and the other describes the detector response.

The choice of comb parameters relies on the shape of resonance structure for echo amplitudes. In the Green’s function approach, the structure originates from a transfer function modulated by a source integral, and narrow resonances correspond to poles living close to the real axis. Although the generic features are known quite well, the exact shape of the resonance structure is subject to various uncertainties~\cite{Conklin:2017lwb, Conklin:2019fcs, Xin:2021zir}. The transfer function relies on the metric of UCOs and the interior damping of gravitational waves. Potentially large deviations from black holes close to the horizon could make the resonance spacing more nonuniform. A strong damping may significantly broaden the resonances. The source integral encodes the initial perturbation dependence, and is hard to specify from the current black hole simulations or observations due to the unknown physics for merger of UCOs. This can bring in large uncertainties for the resonance amplitudes as well as the relative importance of contributions at positive and negative frequencies.
In this study, we consider a uniform comb consisting of triangular teeth to pick out the dominant component of the resonance structure after projection in Eq.~(\ref{eq:response}). Despite all the above uncertainties, it captures the most essential properties of the resonance structure~\cite{Conklin:2017lwb, Holdom:2019bdv} and then deserves to be carefully investigated in our new algorithm.

\begin{table}[h]
\begin{center}
\begin{tabular}{l||l}
\hline\hline
&
\\[-3mm]
Parameters & Descriptions
\\
&
\\[-3.5mm]
\hline
$\Delta f$ & The spacing between two consecutive teeth
\\
$f_0$ & The ratio of comb overall shift to spacing
\\
%\hline
$A_\textrm{comb}$ & The comb amplitude 
\\
%\hline
$f_w$ & The width of the triangular tooth
\\
%\hline
$f_\textrm{min}$ & The frequency band lower boundary 
\\
%\hline
$f_\textrm{max}$ & The frequency band upper boundary
\\
\hline\hline

\end{tabular}
\caption{The six free parameters and the corresponding descriptions for a uniform comb. }
\label{tab:combparameter}
\end{center}
\end{table}

A uniform comb consisting of triangular teeth is fully specified by six parameters, as explained in Table~\ref{tab:combparameter}.
Their physical meanings are intimately related to the generic features of the resonance structure.
The comb spacing $\Delta f$ measures the average spacing between resonances, and it is roughly the inverse of the time delay $t_d$ in Eq.~(\ref{eq:timedelay}) with
\begin{eqnarray}\label{eq:Deltaf}
\frac{\Delta f}{\textrm{Hz}} \approx \frac{\bar{R}}{\eta},\;\;
\bar{R}\approx 583\,\frac{M_\odot}{M} \,\frac{2}{1+(1-\chi^2)^{-\frac{1}{2}}}\,,
\end{eqnarray}
where $\bar{R}$ summarizes dependence on the final object's properties. The overall shift $f_0$ varies within [0,1] by definition, and it is sensitive to the phase shift at the interior reflection boundary. 
The triangular tooth width $f_w$ has to be much smaller than $\Delta f$ to account for the narrow resonances. Meanwhile, it has to be sufficiently wide to incorporate a large number of resonances that are not exactly periodic. The frequency band is applied to cut off the declining spectrum at small and large frequencies. The upper boundary  $f_\textrm{max}$ is around the black hole ringdown frequency~\cite{Berti:2009kk} 
\begin{eqnarray}\label{eq:fRD}
\frac{f_\textrm{RD}}{\textrm{Hz}}=4.944\times 10^4\,\frac{M_\odot}{M} \left[1-0.759(1-\chi)^{0.129}\right]\,,
\end{eqnarray}
for the $l=m=2$ mode.  Above $f_\textrm{RD}$, the wave is no longer trapped inside the cavity and the resonance structure quickly diminishes. The lower end $f_\textrm{min}$ results from the rising noise and declining signal due to a limited frequency resolution when the frequency goes smaller. For spinning UCOs, assuming the same spin dependence as in the case of Kerr black holes, there is a special intermediate frequency~\cite{Conklin:2017lwb}
\begin{eqnarray}\label{eq:fH}
\frac{f_H}{\textrm{Hz}}=3.24\times 10^4\,\frac{M_\odot}{M}\,\frac{\chi}{1+\sqrt{1-\chi^2}}\,,
\end{eqnarray}
at which the transfer function vanishes. Depending on the initial condition or the interior damping, the resonance structure near $f_H$ could be suppressed~\cite{Conklin:2019fcs}. 
From Eqs.~(\ref{eq:Deltaf})--(\ref{eq:fH}), we see that the frequency ratios $f_H/f_\textrm{RD}$ and $f_\textrm{RD}/(\eta\Delta f)$ depend mainly on the dimensionless spin $\chi$. For $\chi$ varying from 0.5 to 0.9, $f_H/f_\textrm{RD}$ and $f_\textrm{RD}/\Delta f$ increase from 0.6 to 0.9 and from $30\eta$ to $60\eta$, respectively. 
If $f_\textrm{min}$ is considerably smaller than  $f_\textrm{RD}$, the number of resonances within the frequency band is of the order of $f_\textrm{RD}/\Delta f$. Thus, for $\eta\sim \mathcal{O}(1)$, the structure does feature dozens of resonances.

Another important parameter that determines the appearance of the resonance structure is the duration $T$ of the strain time series. It determines the number of pulses included for the coherent combination as well as the frequency resolution.  
A short range of data incorporating only the first few pulses can recover the wider resonances at higher frequencies, but loses sensitivity to the low-frequency modes. A  long range of data incorporating many hundreds of pulses can resolve a large number of narrow resonances, but the SNR might not be optimal.  Adjusting $T$ can also produce a more evenly distributed resonance structure with larger overlap with a uniform comb. Thus, a proper choice of $T$ plays an essential role to optimize the search sensitivity. 

The detector response is fully account for by the three parameters $A_{HL}, \phi_{HL,0}$ and $\Delta t_{HL}$ in Eqs.~(\ref{eq:rhomf4}) and (\ref{eq:phiHL}). 
For data within a limited frequency band, $\phi_{HL,0}$ and $\Delta t_{HL}$ have some degeneracy in determining the relative phase. Since the time lag $\Delta t_{HL}$ is very well measured from the  inspiral-merger-ringdown signal, it is better to fix $\Delta t_{HL}$ and vary $\phi_{HL,0}$ around $\pi$ to compensate the mismatch. Once the relative phase is corrected, the amplitude ratio $A_{HL}$ becomes degenerate with the comb amplitude $A_\textrm{comb}$. Here we fix $A_{HL}$ to be one and use $A_\textrm{comb}$ to estimate the average amplitude for two detectors.  

In total, our echo search with a uniform comb requires no more than ten search parameters. The morphology-independent BayesWave algorithm, in contrast, has $9N+4$ parameters for $N$ pulses~\cite{Tsang:2018uie}. To efficiently probe the quasiperiodic  structure, $N$ cannot be too small and the number of parameters increases dramatically with $N$. That is to say, by throwing away the phase information, a significant overlap with a large variety of resonance structures can be achieved by using combs without introducing too many parameters.

We use BILBY, a user-friendly Bayesian inference library in PYTHON, to analyze data~\cite{Ashton:2018jfp}. Its modular structure can be easily adapted to incorporate the phase-marginalized likelihood and the comb model for our search. To explore the complicated likelihood distributions in the high-dimensional parameter space, we employ the nested sampling algorithm~\cite{2004AIPC..735..395S, 10.1214/06-BA127} with the default DYNESTY sampler~\cite{2020MNRAS.493.3132S}. 
Since the search target incorporates a large number of narrow resonances, a great fine-tuning of the comb spacing $\Delta f$ is required to achieve a significant overlap between the comb and the signal. A proper choice of the sampler settings is then essential for an efficient detection. We choose $n_\textrm{live}=1000$, walks$=100$, $n_\textrm{act}=10$ and maxmcmc$=10000$ in this study.\footnote{For the stochastic sampler in BILBY, $n_\textrm{live}$ is the number of live points used in the nested sampling, ``walks'' is the number of walks to take each time a new point was proposed, ``maxmcmc'' is the maximum number of walks to use, and $n_\textrm{act}$ is the number of autocorrelation times to use before accepting a point~\cite{Ashton:2018jfp}.} In addition, we run $n_\textrm{run}=3$ analyses in parallel for each dataset to avoid the bias of a single run on a multimodal likelihood distribution~\cite{Romero-Shaw:2020owr}. The posterior samples from parallel runs are combined with the weight $w_i=\mathcal{B}_i/\sum_i \mathcal{B}_i$, and the combined Bayes factor  is $\mathcal{B}=\sum_i \mathcal{B}_i/n_\textrm{run}$.

%%%%%%%%%%%%%%%%%%%%%%%%%%%%%%%%%%%%%%%%%%%%%%%%

\section{Results and Discussions}
\label{sec:result}

It is necessary to validate the performance of the search algorithm before doing searches on real data. In the following, we first consider signal injections in stationary Gaussian noise.
Then we move to analysis on real gravitational wave strain data, with special treatments for data quality and parameter choices.

\subsection{Analysis in the presence of stationary Gaussian noise}\label{sec:Gaussian}

As a first try, we directly inject uniform combs in stationary Gaussian noise to better understand the behavior of the phase-marginalized likelihood. Combs with 30-50 teeth are considered to mimic the echo signal. We use the normalized likelihood function in Eq.~(\ref{eq:logL2}) with the SNRs given in Eq.~(\ref{eq:rhomf2p}) to evaluate the log Bayes factor for the comb versus noise models.
Among all comb parameters in Table~\ref{tab:combparameter}, $\Delta f, f_0$ and $A_\textrm{comb}$ are considered as search parameters, and the rest are fixed as the injected values. We choose uniform prior distributions for the search parameters, with the prior ranges: $\Delta f\in[ 1,10]\,$Hz, $f_0\in [0,1]$ and $A_\textrm{comb}\in [0,0.5]$\,Hz$^{-1}$. For the amplitude, a uniform prior works because the algorithm loses sensitivity to signals significantly below noise, and the upper bound (with the normalized value $8.9$) is set to be well above the noise fluctuation.\footnote{The exact upper bound of the comb amplitude does not really matter, since the log-likelihood drops quickly for increasing amplitude.}

\begin{figure}[!h]
  \centering%
{ \includegraphics[width=9.4cm]{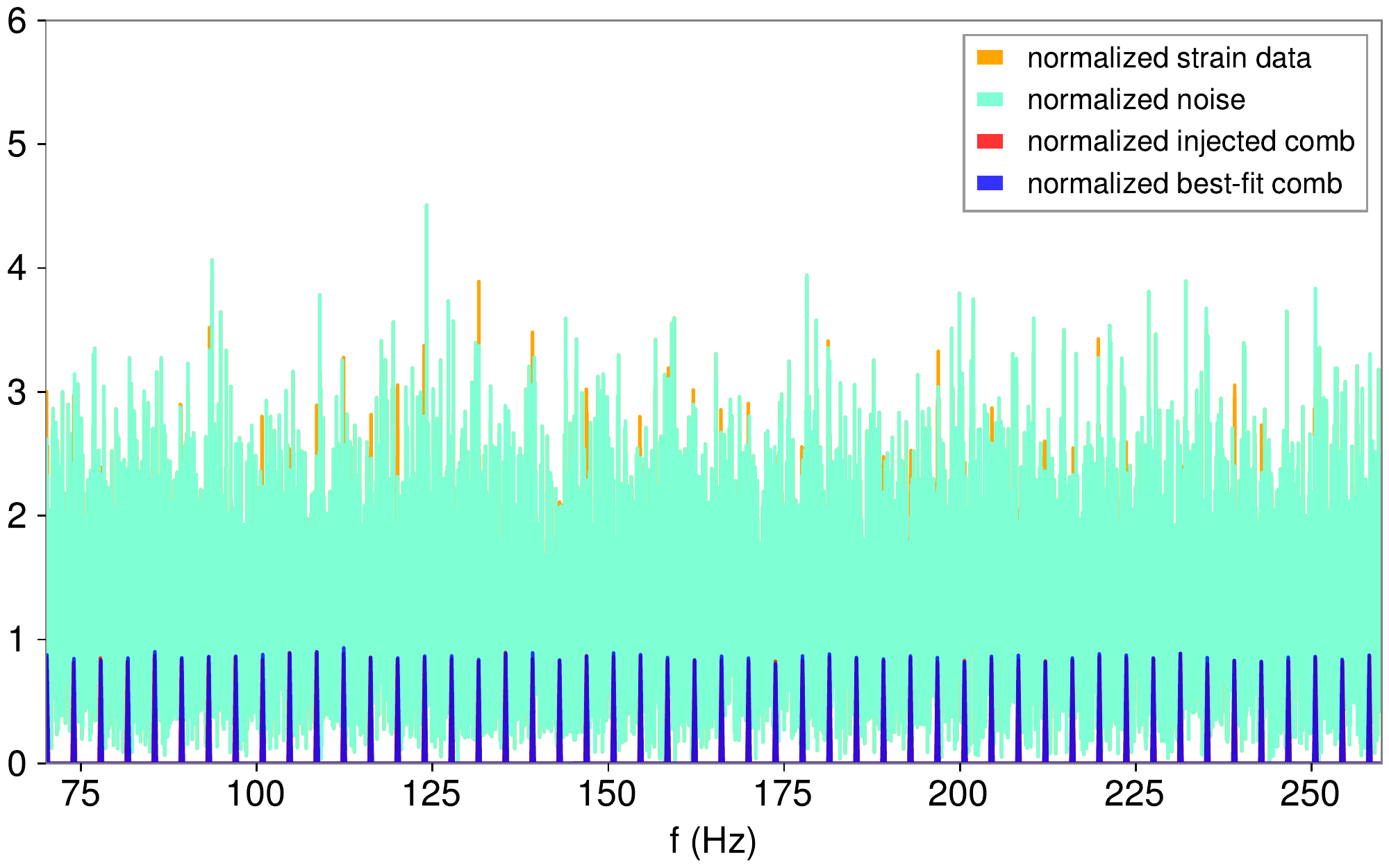}}\;
{ \includegraphics[width=6cm]{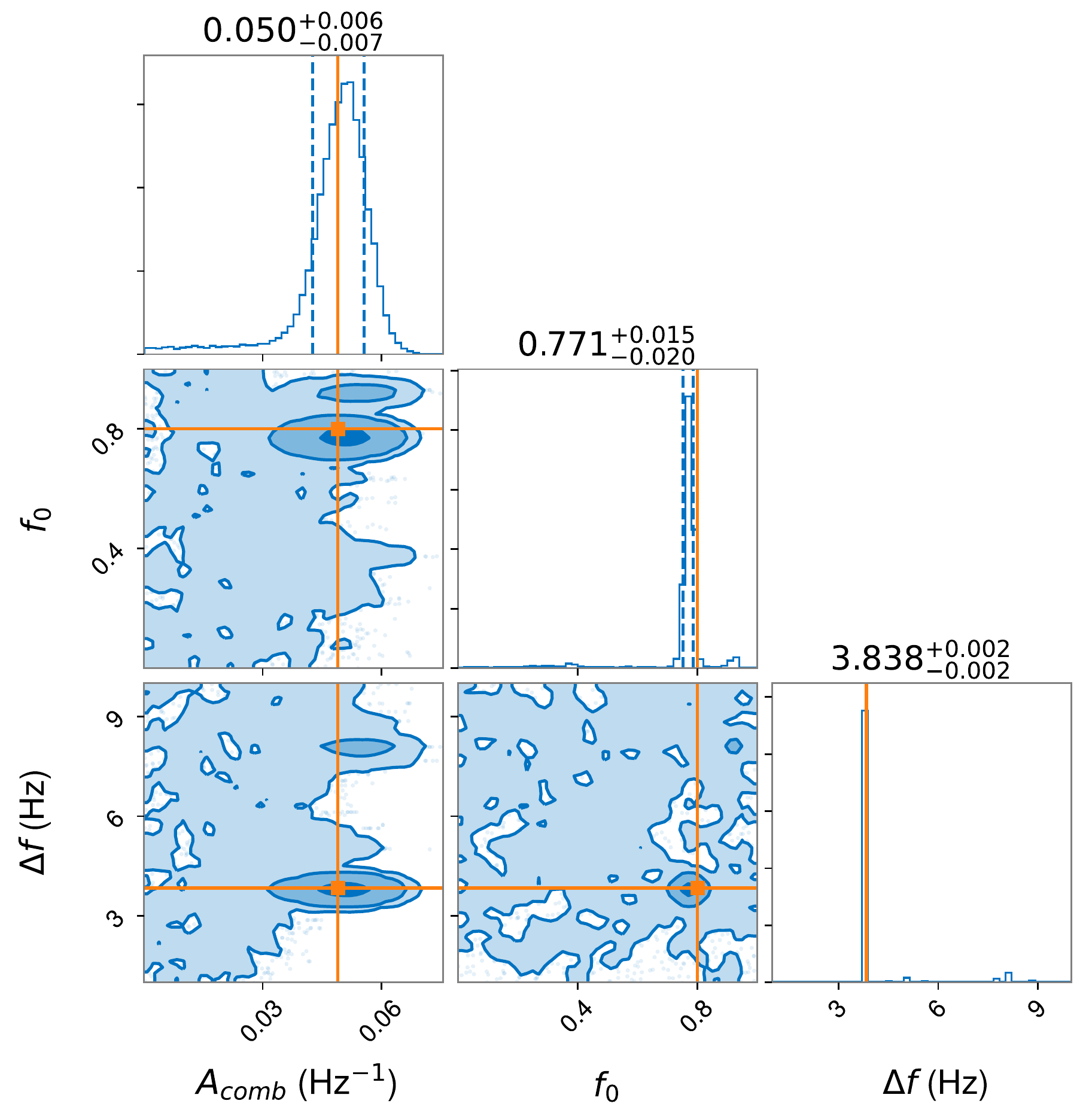}}
\caption{\label{fig:bestfit1} The search results for an injected comb with the network optimal $\textrm{SNR}\approx 16.2$ and $\ln\mathcal{B}\approx 0.24$. Left: the absolute value of the normalized data in the frequency space. The orange (cyan) line denotes the combined strain (noise) data $\tilde{d}_{IJ}$  in Eq.~(\ref{eq:rhomf2p}) for two detectors. The excess of orange comes from the injected comb (the red line), which lines up well with the normalized best-fit comb $\tilde{h}$ (the blue line) in Eq.~(\ref{eq:rhomf2p}), as drawn from the best estimates of the comb parameters (the median or maximum of the posterior distribution). Right: the corner plot for the sampled posterior distributions in blue and the injected values in orange. The dashed vertical lines denote the 68\% credible interval. The diagonal panels show the estimated 1D marginal posterior probability distribution for each parameter, and the off-diagonal panels show the joint distribution for each pair of parameters. }
\end{figure}

Figure~\ref{fig:bestfit1} displays the search results for an injected comb with the normalized comb amplitude around one~\cite{Hunter:2007, corner}.
From the left panel, we can see that the small excess of the strain data due to the injected comb is quite comparable in magnitude to the noise fluctuation. This gives some idea about the smallness of a signal to be successfully recovered by using the phase-marginalized likelihood. The three comb parameters are precisely measured. 
The spacing $\Delta f$, in particular, is determined extremely well, with the error even smaller than $0.1\%$ of the prior range. This is not surprising because the search target incorporates a larger number of teeth, and a small shift of $\Delta f$ destroying the perfect match can significantly reduce the likelihood. The other two parameters are less precisely estimated, and the errors are of the order of $1\%$ of the prior ranges with no extra suppression from the number of teeth. 
The inferred comb amplitude is slightly larger than the injected value due to the nonzero contribution from noise after taking the absolute value. The mismatch becomes smaller with increasing amplitude of the injected signal.

\begin{figure}[!h]
  \centering%
{ \includegraphics[width=8.8cm]{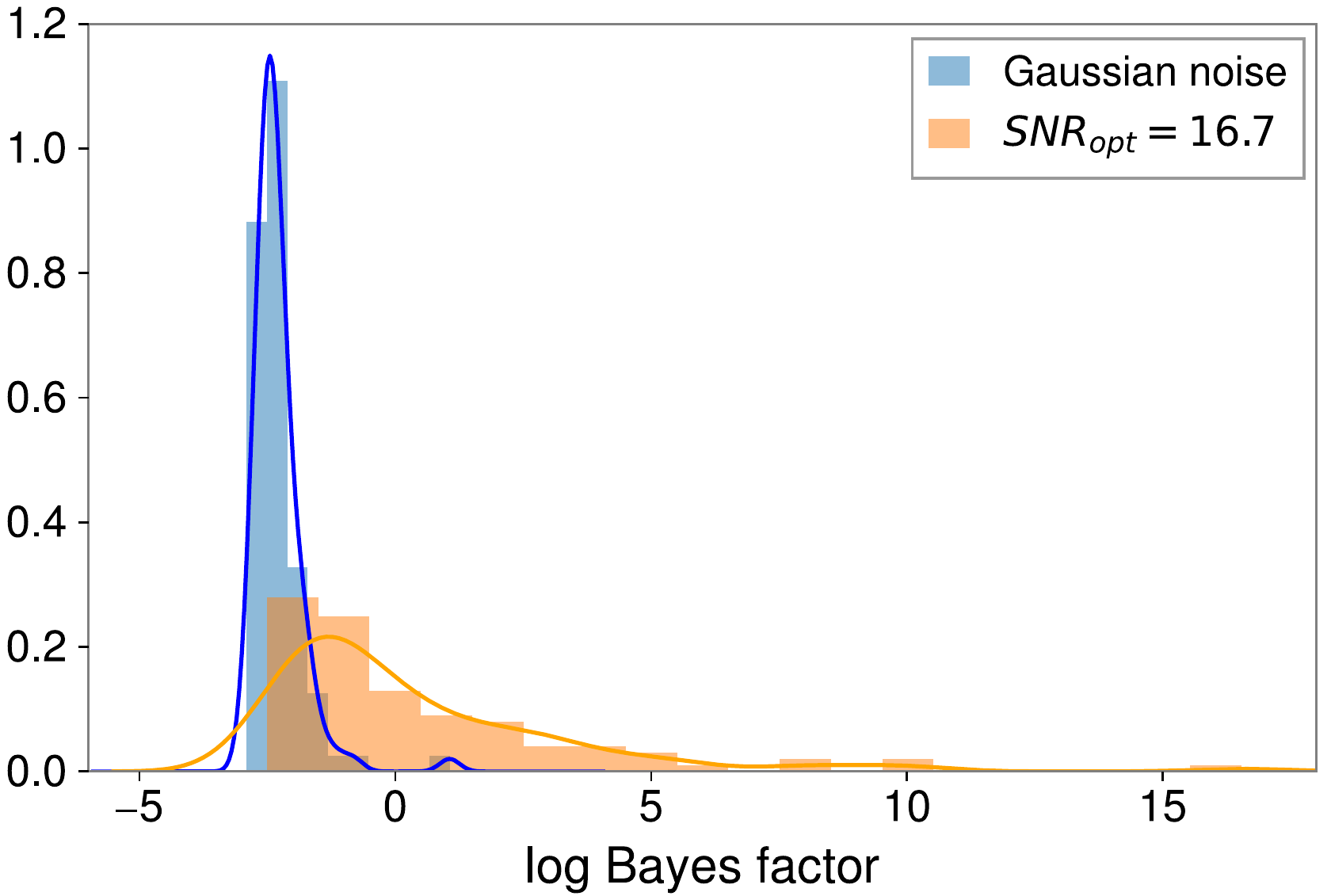}}\;\;
{ \includegraphics[width=6.6cm]{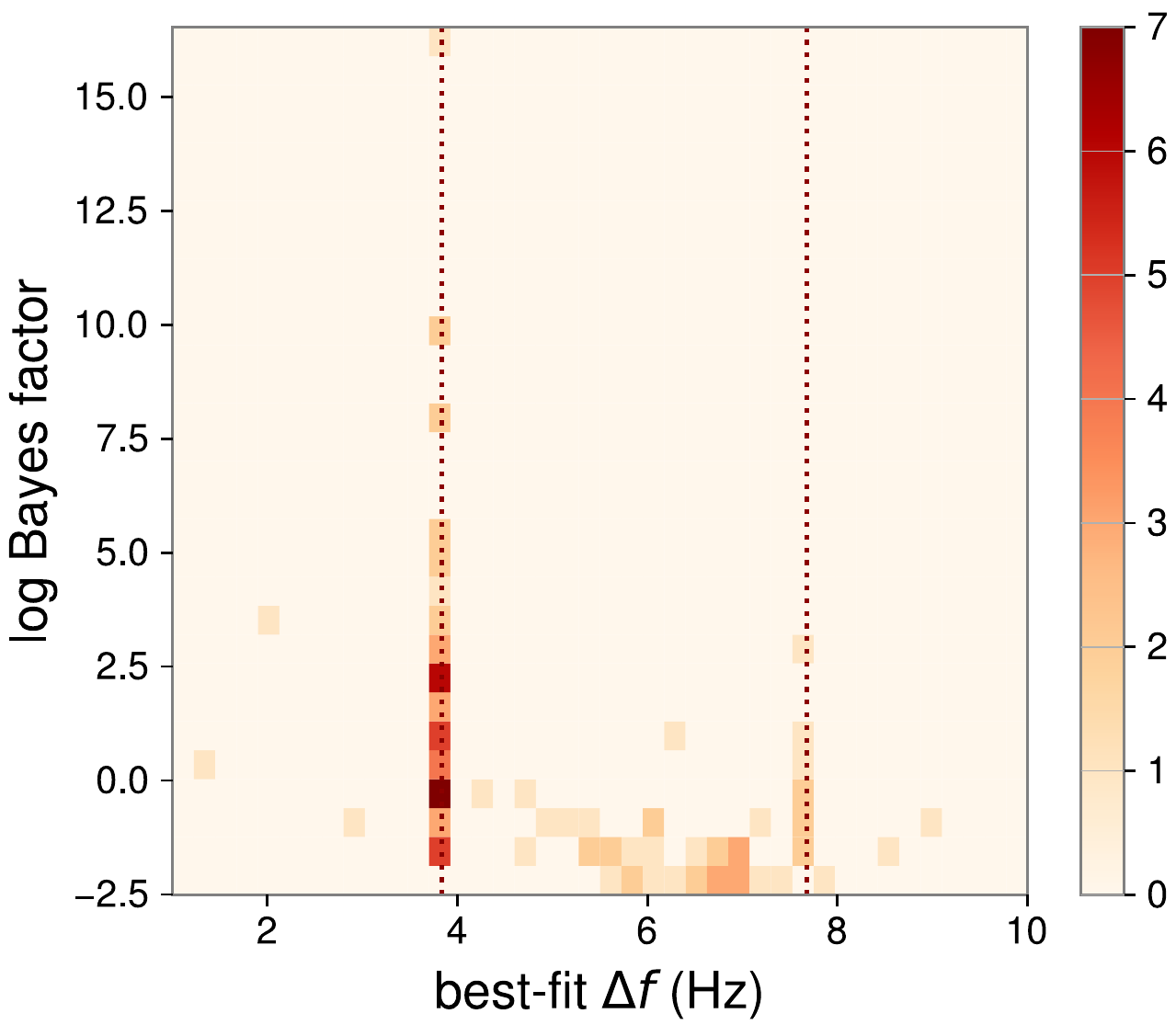}}
\caption{\label{fig:logB_comb} Left: the distributions of the combined log Bayes factor $\ln\mathcal{B}$ for the comb versus noise models with 100 samples of stationary Gaussian noise. The histograms in blue and in orange show the normalized background distribution and signal distribution, respectively, by analyzing Gaussian noise without and with  injection of the comb signal in Fig.~\ref{fig:bestfit1}. The solid lines are smoothened distributions using Gaussian kernel density estimations. Right: the 2D distributions of the best-fit comb spacing versus the log Bayes factor for the signal samples. The two vertical dotted lines denote the injected value and twice the injected value of spacing.}
\end{figure}

The distributions of the combined log Bayes factor for the comb versus noise models are displayed in Fig.~\ref{fig:logB_comb}. The background distribution is quite narrow, and the values of the log Bayes factor are mostly below zero. The signal distribution, in contrast, is wide with a long tail extending to large values. 
When the likelihood function features a large peak in the parameter space, the log Bayes factor can be well approximated as the sum of the maximum log-likelihood ratio and the Occam penalty factor\;$\sim \ln \Delta V/V$, where $\Delta V$ is the likelihood spread around the maximum and $V$ is the total parameter volume~\cite{Romano:2016dpx}. The Occam factor is always negative and then penalizes models with a larger parameter space. For the comb model, the requirement of a fine-tuning of the comb spacing leads to a strong Occam penalty, indicating more conservative criteria for model selection by using the log Bayes factor. 
With the smoothened probability distributions, we find the detection probability of this injected comb signal around 76\% for a false alarm probability equal to 10\%.\footnote{A different choice of PSD settings may shift the overall values or change the shapes of the log Bayes factor distributions, but the effects on the detection probability are minor.} 
For a weaker signal, the detection probability may not scale with the optimal SNR due to the noncoherent combination of different frequency bins in the phase-marginalized likelihood. 
It is detectable only if the strain data that dominate the likelihood take the large $x$ expansion in Eq.~(\ref{eq:logI0}). 
As another feature for the comb model search, the inferred spacing could be multiples of the injected value due to the periodic structure of the comb. For the search considered here, we see twice the injected value recovered for a small fraction of samples in the right panel of Fig.~\ref{fig:logB_comb}.

%%%%%%%%%%%%%%%%%%%%%%%%%%%%%%%

We next validate the algorithm by considering injections of more realistic echo waveforms produced from some toy models. The search sensitivity is expected to be degraded by the mismatch between the resonance structure and a uniform comb. To minimize their difference, we vary the duration $T$ of the time series data to adjust the shape of the resonance structure and then apply a frequency band specified by $f_\textrm{min}$ and $f_\textrm{max}$ to select the frequency region of interest. 
For the sake of a reasonable run time, the duration $T$ is not considered as a search parameter. Instead, we fix $T$ to some value for each search and include $ f_\textrm{min}$ and $f_\textrm{max}$ as two additional search parameters. A preliminary study shows a weak dependence of the search on the tooth width, and so we fix the width value in this study. A more thorough exploration is left for future work.

\begin{figure}[!h]
  \centering%
{ \includegraphics[width=8cm]{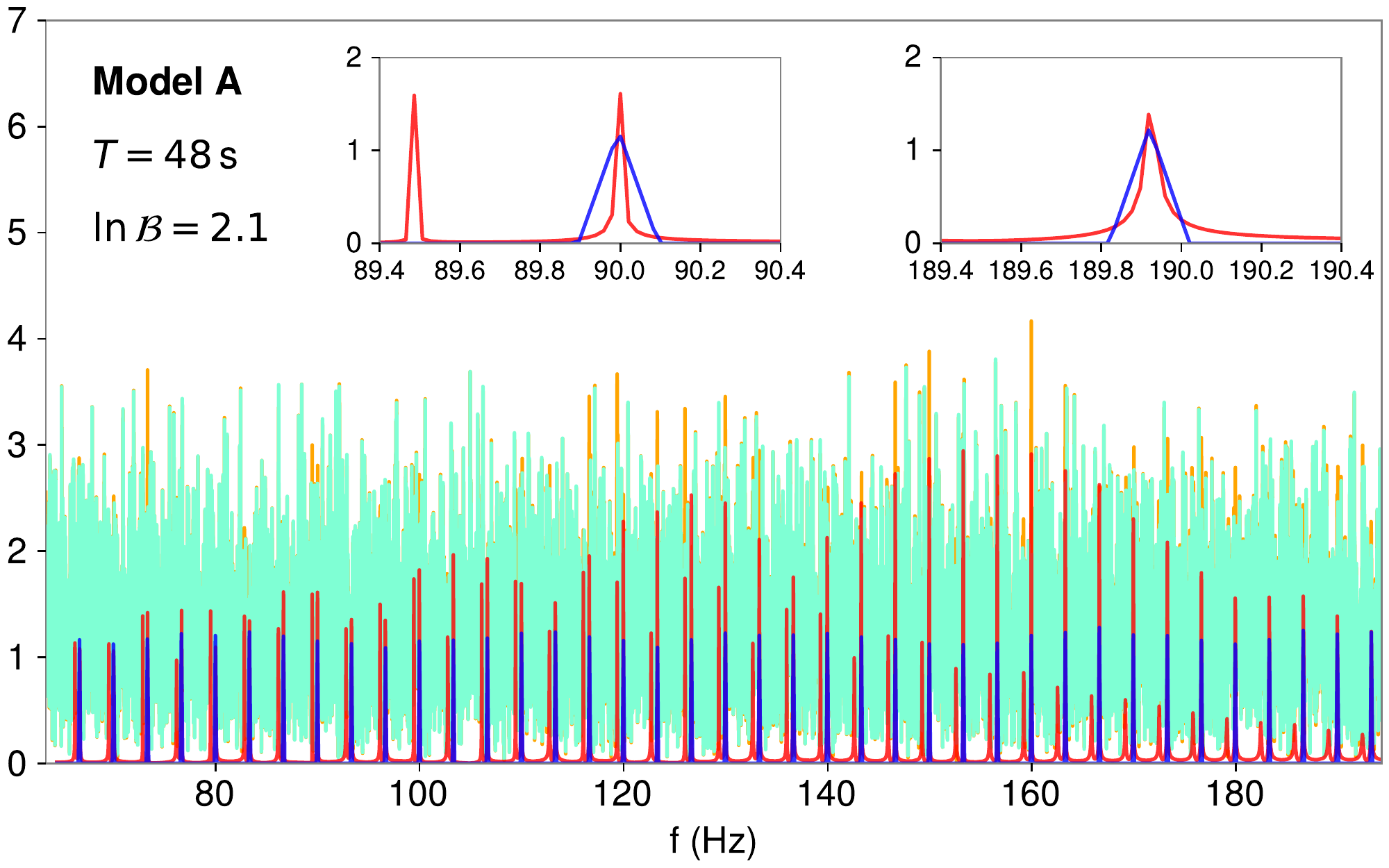}}\;\;
{ \includegraphics[width=8cm]{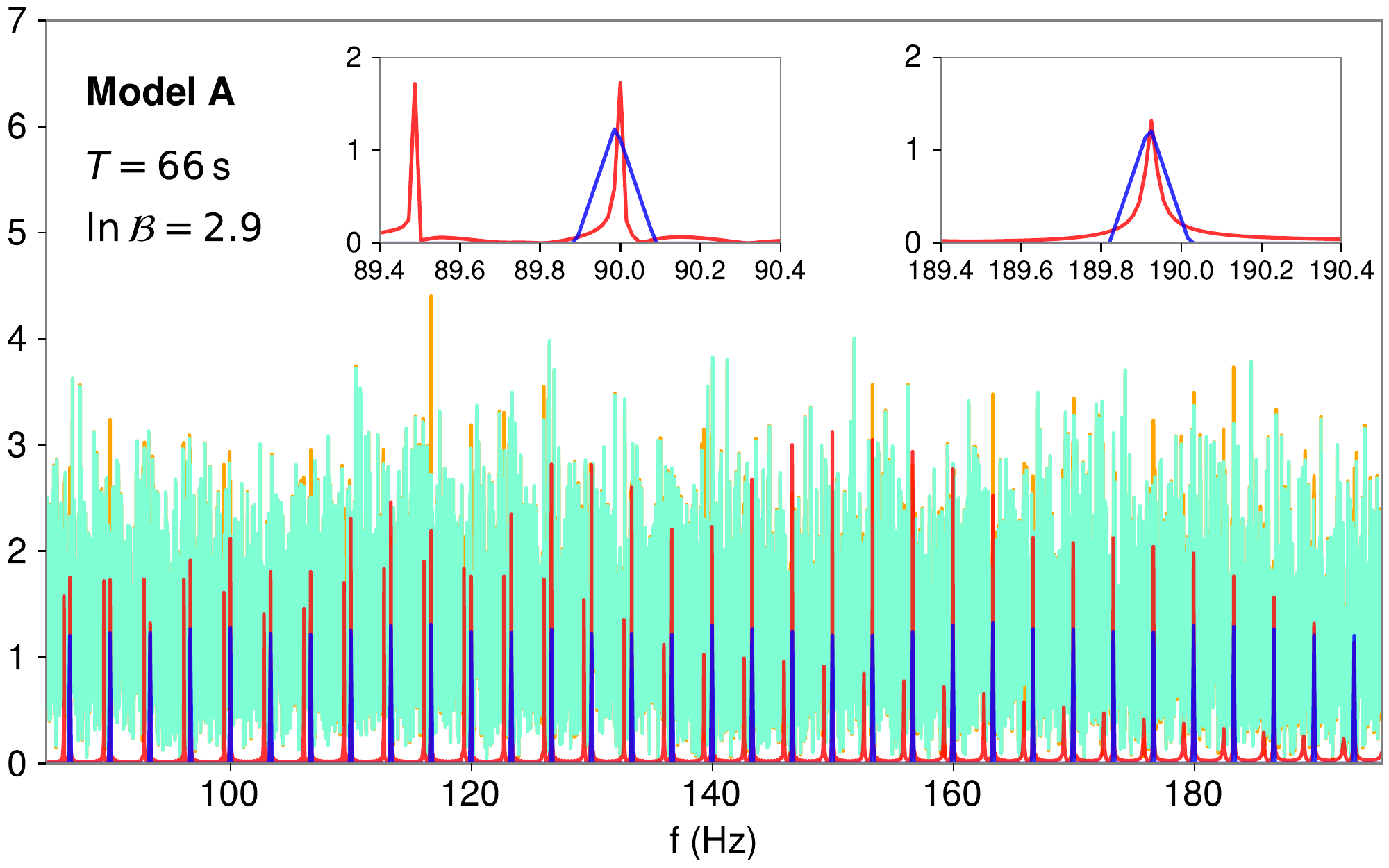}}
{ \includegraphics[width=8cm]{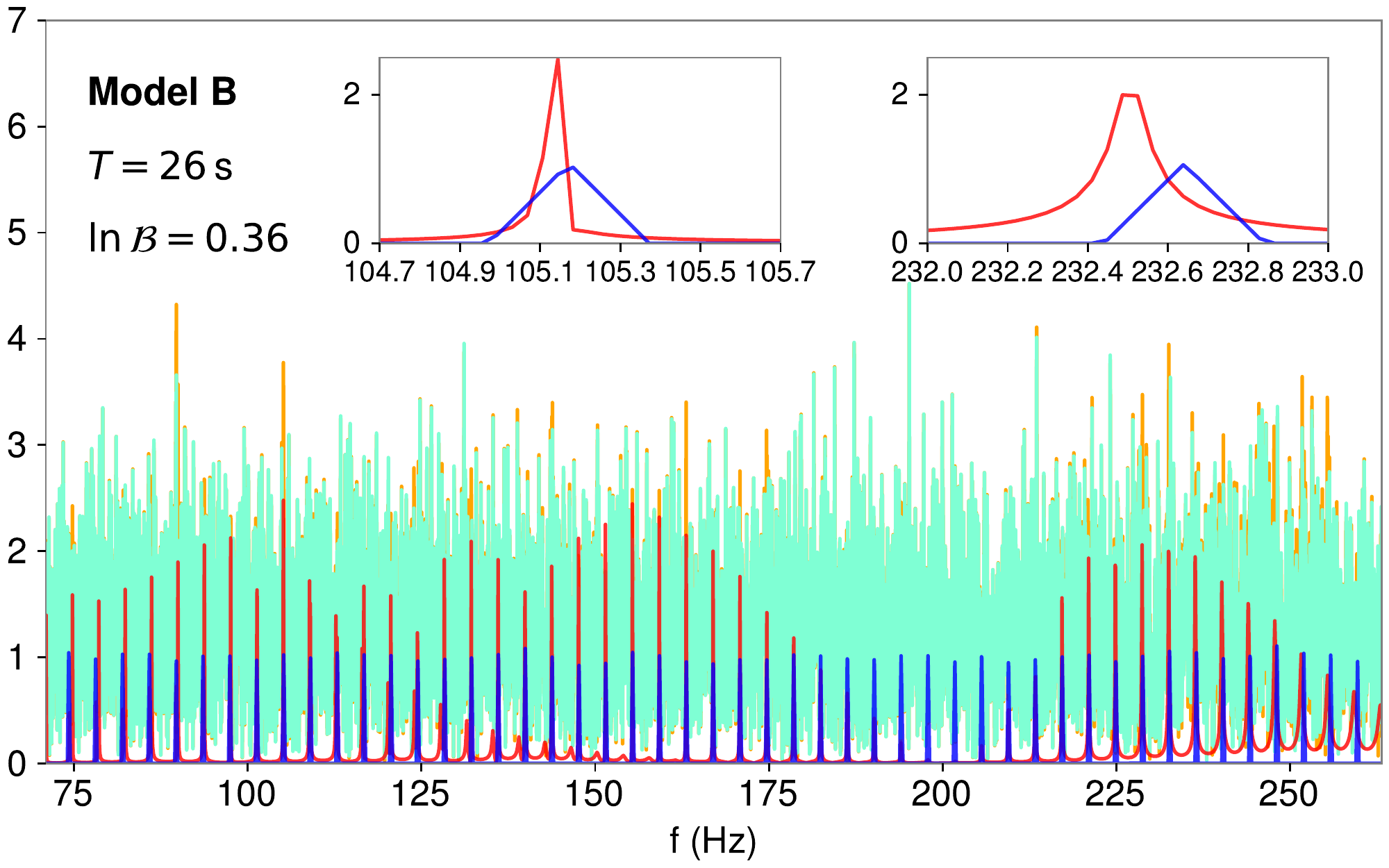}}\;\;
{ \includegraphics[width=8cm]{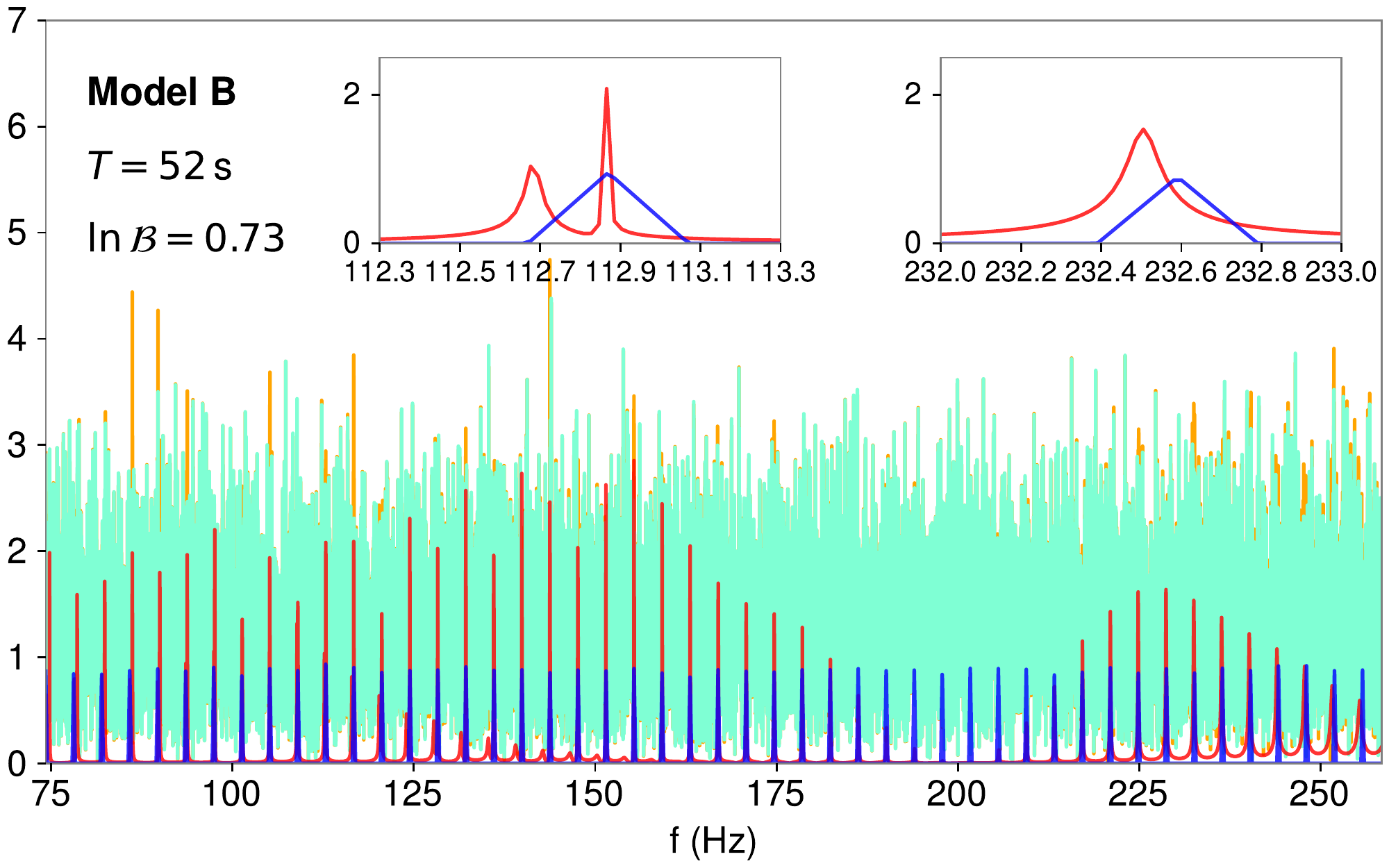}}
\caption{\label{fig:bestfit1G} The absolute value of the normalized data with an injected echo signal from model A and model B with two different time durations $T$. In each panel, as in Fig.~\ref{fig:bestfit1}, the orange (cyan) line denotes the combined strain (noise) data for two detectors, and the blue line is the normalized best-fit comb $\tilde{h}$. The red line denotes the injected echoes from toy models with the network optimal SNRs around 18, and the average spacing $3.33$ and $3.84\,$Hz for model A and B, respectively.  The two insets show the detailed comparison between one resonance and the corresponding tooth at low and high frequencies, respectively.}
\end{figure}

For comparison, we consider two toy models for the echo waveform that differ mainly by assumptions on the initial condition for echoes. 
Model A, the analytical waveform in Fig.~\ref{fig:echo2}, relates the initial perturbation directly to the ringdown waveform for the Kerr black hole~\cite{Maggio:2019zyv}, and so the frequency content of echoes remains the same as that for the black hole ringdown. Model B, on the other hand, considers echoes produced dominantly by disturbances originating in the core of the newly forming UCO~\cite{Conklin:2017lwb}, which can be modeled by an initial pulse starting inside  the light ring and moving toward the barrier \cite{Conklin:2019fcs}. The main difference for our search is the shape of the resonance structure. Below the ringdown frequency $f_\textrm{RD}$, the pattern is more uniform for model A, while it is suppressed around $f_H$ that the transfer function vanishes for model B. For both cases, an almost perfect reflectivity of the interior is assumed, and the resonances are almost the narrowest in theory. 
Given the aforementioned theoretical uncertainties, these models are used here to demonstrate the capability of our algorithm in recovering resonance structures of different shapes, and the echo amplitude is set as a free parameter.

Figure~\ref{fig:bestfit1G} shows search results of examples of injected echoes from the two toy models. Their echo amplitudes are chosen to have the network optimal SNR comparable to that of the injected comb considered before. Since the frequency band of model B is similar, the average height of resonances is comparable to the comb amplitude in Fig.~\ref{fig:bestfit1}. Model A has a smaller frequency band, and then features higher resonances on average.  For each case, two different time durations $T$ are considered, corresponding to the number of pulses ranging from 100 to 200. The tooth width is fixed around the minimum value for a uniform comb to capture dozens of resonances, and  for model A and B we find $f_w=0.2$ and $0.4\,$Hz respectively. The prior ranges for $\Delta f, f_0$ and $A_\textrm{comb}$ remain the same as in the case of injected comb study. For the frequency band, we take uniform priors: $f_\textrm{min}\in[50,100]$\,Hz for both models, and $f_\textrm{max}\in [150,250]$\,Hz for model A ([200,300]\,Hz for model B). As expected, the low-frequency part of the injected signal (the red line) becomes more important with increasing time duration. For model B, in particular, a quite evenly distributed structure can be obtained for $T/t_d\approx 100$.
%$T/t_d\approx 100$. 

For a realistic resonance structure, the inferred comb spacing measures the average spacing between resonances. It remains very well determined, with the error of the order of $0.01\%$ and $0.1\%$ of the prior range for model A and B, providing an excellent estimation for the time delay with $t_d\approx 1/\Delta f$. 
The inferred comb amplitude measures the average heights of resonances to have comparable SNRs for the best-fit comb and for the signal. The average comb amplitude for model A is higher than that in Fig.~\ref{fig:bestfit1}, indicating a stronger signal with better search sensitivities. 
For model B, the resonances away from $f_H$ are significantly higher than the best-fit comb to compensate the reduced power around $f_H$. The frequency band is determined up to the order of $10\%$ of the prior range, indicating the important roles played by the two new parameters. From the insets, we see that the triangular teeth are wider than most of the resonances, and this helps to increase the overlap between a uniform comb and a nonuniform resonance structure.

\begin{figure}[!h]
  \centering%
{ \includegraphics[width=16cm]{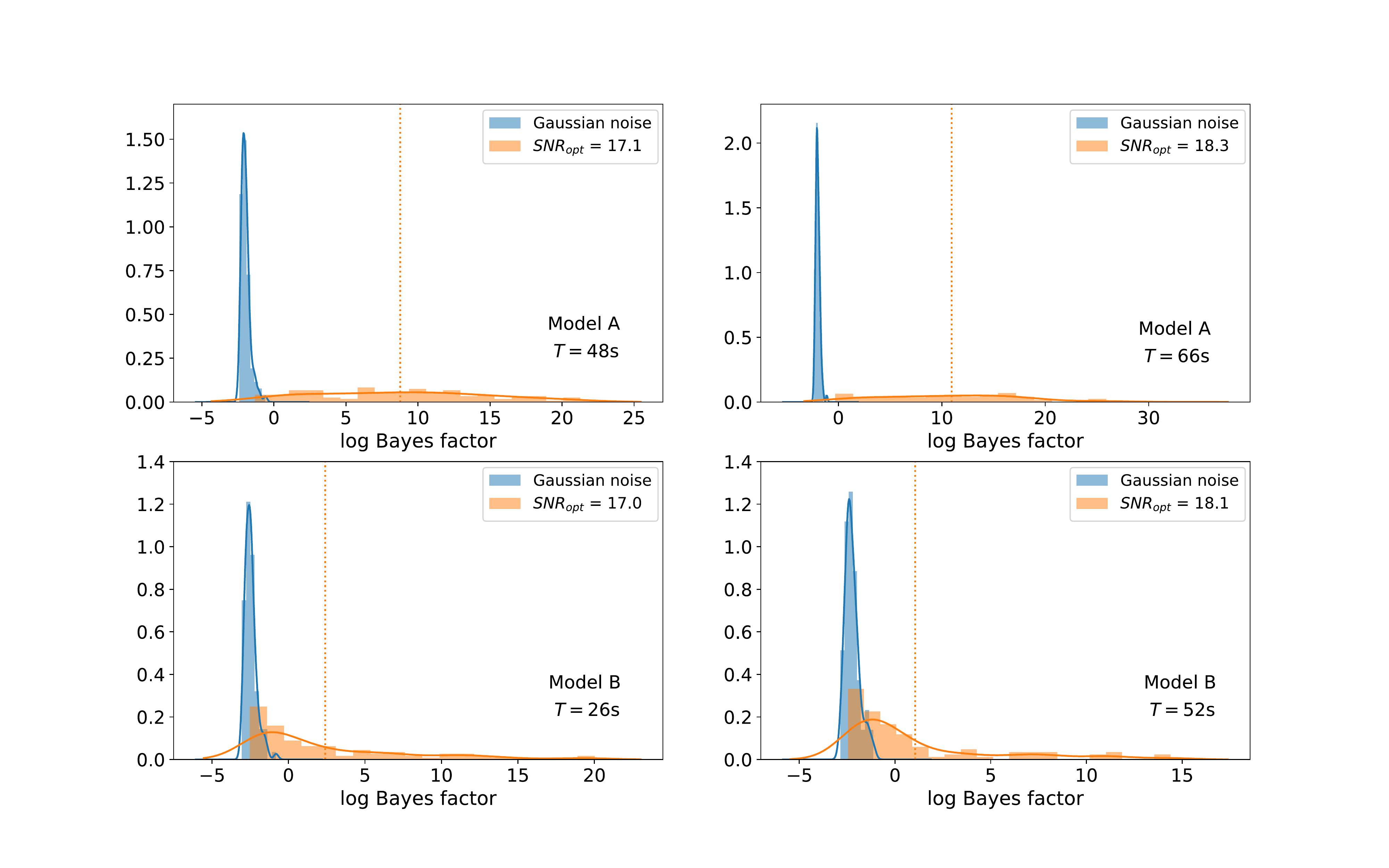}}
\caption{\label{fig:logB_echo} The distributions of the combined log Bayes factor $\ln\mathcal{B}$ for different injected echo signals in Fig.~\ref{fig:bestfit1G}. The histograms in blue and in orange show the normalized background and signal distributions, respectively. The orange dotted line denotes the average value of the log Bayes factor.}
\end{figure}

Figure~\ref{fig:logB_echo} presents the log Bayes factor distributions for injected echo signals from two models. 
The log Bayes factor distributions for model B are similar to that for the injected comb signal in Fig.~\ref{fig:logB_comb} since 
their average SNRs and inferred comb amplitudes are comparable. The detection probability of the injected echo signal (for a false alarm probability equal to 10\%) is 79\% (73\%) for $T=26$ (52)\,s. For model A, the signal distribution move considerably to the right due to higher inferred comb amplitude on average. The detection probability of the injected echo signal is then much larger, i.e., 96\% (97\%) for $T=48.6$ (66.3)\,s. Thus, instead of the SNR, the average height of resonances plays a more important role in determining the search sensitivity due to the noncoherent combination of frequency data in the phase-marginalized likelihood.

It is useful to make comparison with searches targeting on the quasiperiodic  signal in the time domain.  
Since different methods are tested by a variety of signal models, the sensitivities cannot be directly compared. Rather, we focus on their complementarity in probing the large morphology space of echoes here. 
If the real signal damps fast, with the SNR dominated by a few early pulses of large amplitudes and clear shapes, 
the morphology-independent search based on the BayesWave algorithm shall work better, and the network optimal SNR for the smallest detectable signal is found $\sim12$ for the initial five pulses~\cite{Tsang:2018uie}. 
Otherwise, our resonance search is better for a signal dominated by a large number of slowly damped small pulses at late time. For the injected signal we considered above, the SNRs of the initial five pulses and of the rest are $\sim 5$ and $\sim 17$ for model A, and $\sim 10$ and $\sim 14$ for model B.

Once an echo signal is identified, another important difference comes down to the precision of parameter estimations for the time delay $t_d$. For the resonance search, the time delay is inferred from the periodicity of resonance locations, and the precision is strongly enhanced by the number of resonances of comparable sizes in the spectrum. From toy models considered in Fig.~\ref{fig:bestfit1G}, we see that it is quite generic to have dozens of such resonances with a proper choice of $T$. 
On the other hand,  the time domain searches look for a smaller number of pulses that damp constantly with time, and the likelihood  is less sensitive to variations of the pulse separation, leading to a worse precision for $t_d$. As for an explicit comparison, the time delay precision of the order of $0.1\,$\% or below is achieved in our search with the network optimal SNR below 20, while the $t_d$ precision is found to be of the order of $1\,$\%~\cite{Tsang:2018uie}  and $0.1\,$\%~\cite{Lo:2018sep} for SNR$\,\sim25$ and 64 with two time domain search methods.

In summary, our search algorithm with a phase-marginalized likelihood and a uniform comb is able to detect the resonance structure of echoes  in a rather model-independent way. 
Variation of the time duration $T$ helps to adjust the shape of the resonance structure, and then enhances the detection probability. The inferred comb spacing and overall shift provide a precise estimation of the resonance locations. Other parameters such as the comb amplitude and the frequency band do not have clear definitions from the echo signal, but they tell roughly the average height of the resonances and the frequency region of interest.

\subsection{Analysis of Advanced LIGO data}

The real detector noise is neither stationary nor Gaussian. For our search of narrow resonance structure, the main non-Gaussian artifacts of concern are narrow spectral lines due to instrumental or environmental disturbances~\cite{Covas:2018oik,Davis:2021ecd}. Among all reported lines, some are clearly identified as instrumental due to correlations with environmental monitoring channels, while many remain unknown in origin. As in the case of SGWB searches, lines coherent between two detectors cause the major problem. Noncoherent lines may also contribute in our search if they fall in different teeth of one comb. 

To demonstrate the algorithm performance, we focus on the first observing run (O1) of Advanced LIGO in this study. The O1 strain data are known to be polluted by a large number of instrumental lines and, thus, provides a good place to validate our algorithm. These lines can easily produce outliers with their large contributions to the likelihood even in a small number of frequency bins. To address this issue, we notch out large instrumental lines above some threshold by assuming that the frequency bins around the lines are badly contaminated and unusable in the search. More explicitly, we apply notch filters to sufficiently large lines that are clearly identified as instrumental, and then remove the frequency bins influenced by notching entirely from the analysis. This is a conservative approach that may reduce the signal sensitivity. But, as we will show below, among all spectral lines reported by the collaboration~\cite{O1line}, only a small number of the prominent peaks have to be considered. Extremely narrow combs with unknown origin are irrelevant due to the limited frequency resolution. 
Since the search target incorporates a large number of resonances, notching at most influences a few of them, and the effects on the search sensitivity are minor. 
Meanwhile, the window functions commonly used to eliminate edge effects in discrete Fourier transform of a finite range of strain time series are found to bring in artifacts and degrade the search sensitivity. A better strategy is to whiten a longer segment of data, and then discrete Fourier transform data in the middle~\cite{Holdom:2019bdv}. We employ the second method in this study.

\begin{table}[h]
\begin{center}
\begin{tabular}{l||l}
\hline\hline
&
\\[-3mm]
Parameters & Priors and fixed (scan) values
\\
&
\\[-3.5mm]
\hline
$\Delta f$ & uniform in $[\bar{R}_{\textrm{min}}/2$, $\bar{R}_{\textrm{max}}/1]\,$Hz
\\
$f_0$ & uniform in $[0,1]$
\\
%\hline
$A_\textrm{comb}$ & uniform in $[10^{-25}\,\textrm{Hz}^{-1}, 5\langle\tilde{P}\rangle^{1/2}]$
\\
%\hline
$f_\textrm{min}$ & uniform in $[f_\textrm{cut}, f_H-\frac{1}{4}(f_H-f_\textrm{cut})]$
\\
%\hline
$f_\textrm{max}$ & uniform in $[f_H, 1.1f_\textrm{RD}]$
\\
%\hline
$\phi_{HL,0}$ & uniform in $[\pi/2,3\pi/2]$
\\
\hline
$f_w$ & $11/T$
\\
%\hline
$A_{HL}$ & $1$
\\
%\hline
$\Delta t_{HL}$ & $\Delta t_{HL,0}$
\\
\hline
$T$ & $T_\textrm{min}+\frac{1}{3}n(T_\textrm{max}-T_\textrm{min})$, $n=0...3$ 
\\
%\hline
\hline\hline

\end{tabular}
\caption{The parameter settings for the real data search. $\Delta f$, $f_0$, $A_\textrm{comb}$, $f_\textrm{min}$, $f_\textrm{max}$, $\phi_{HL,0}$ are search parameters and the priors are given. $\bar{R}_{\textrm{min} (\textrm{max})}$ denotes the minimum (maximum) value of $\bar{R}$ in (\ref{eq:Deltaf}) given the 90\% interval of $M, \chi$ from analysis of the inspiral-merger-ringdown signal. The upper bound of $A_\textrm{comb}$ is defined in terms of the average normalized PSD  $\langle\tilde{P}\rangle$  for two detectors.  $ f_\textrm{RD}, f_H$ are given in (\ref{eq:fRD}), (\ref{eq:fH}), with the best-fit values of  $M, \chi$. $f_\textrm{cut}=\max(50\textrm{Hz}, f_H-30 \bar{R})$ defines the lower end of $f_\textrm{min}$. The frequency band satisfies an additional constraint $f_\textrm{max}-f_\textrm{min}>10\Delta f$. $f_w, A_{HL}, t_{HL}$ are fixed parameters. The value of $f_w$ is  in accordance with the previous study~\cite{Holdom:2019bdv}. $t_{HL}$ is fixed as the best-fit arrival time lag $t_{HL,0}$ for two detectors. Variation of the time duration $T$ is implemented manually by scanning over four benchmark values in between $T_\textrm{min}=100/\bar{R}_{\textrm{max}}\,$s and $T_\textrm{max}=400/\bar{R}_{\textrm{min}}\,$s. (Note that $M$ here is the detector frame mass.)}
\label{tab:parameter}
\end{center}
\end{table}

With these spectral treatments for data quality, we substitute the strain data into the normalized likelihood function in Eq.~(\ref{eq:logL2}), and use expressions in Eq.~(\ref{eq:rhomf4}) with the detector response taken into account. For a confirmed gravitational wave event, the parameter choice for the postmerger echo search is determined by the inferred properties of the final object from analysis of the inspiral-merger-ringdown signal. Our parameter settings for the real data search are summarized in Table~\ref{tab:parameter}. 
There are six search parameters in total, with uniform priors given in the table. The comb spacing $\Delta f$ varies in a range corresponding to a coordinate and a proper Planck distance deviation outside of the would-be horizon. The amplitude $A_\textrm{comb}$ is set to be below a few times of the average noise amplitude. A resonance structure higher than this on average shall be visible. $f_\textrm{min}$ is considered to be above $50\,$Hz to avoid large noise PSD and to incorporate no more than 30 resonances for the limited choice of $T$. To suppress a large contribution from a few noise spikes, the frequency band is set to include a sufficiently large number of resonances. 
The width $f_w$ and  the response-related parameters $ A_{HL}$ and $\Delta t_{HL}$ are fixed either due to weak dependence or due to degeneracy with other parameters. 
For the time duration $T$, four different values with $T \Delta f \in[100, 200]$ are considered to increase the detection probability.  
Note that the optimal choices for $f_w$ and $T \Delta f$ may depend strongly on the echo signal models. Here, we choose their values to be in accordance with those in Ref.~\cite{Holdom:2019bdv} as for a cross-check of the reported signals. For other comb parameters, the inferred values of the reported signals are also covered by the choices in Table~\ref{tab:parameter}.

For a confirmed event, we perform the echo search as follows. First, we do background estimation with  the strain data publicly available in Gravitational Wave Open Science Center~\cite{LIGOScientific:2019lzm}  by using the time slides method. To avoid contamination from either echoes or early inspiral signals, we use the stretches of data preceding merger, in particular, data 250\,s before the GPS time of the merger~\cite{Tsang:2019zra}.
For a given time duration $T$, we define 25 subintervals of duration $T$ for each detector, and then consider a pair of subintervals (one from H1 and one from L1) with nonzero time shift as one noise sample for the analysis.\footnote{The value of the time shift is $T$, and is much larger than the maximum light travel time between two detectors.} For most of the confirmed events and $T$ in Table~\ref{tab:parameter}, it is enough to consider the 4096\,s segment of strain data around merger. The noise does not vary much within such a short period of time and, thus, it is legitimate to use detector noise away from the signal in the segment to estimate backgrounds.
The influence of instrumental spectral lines is closely investigated with the background study. Then, we perform the search on data right after merger of the same duration, and report the $p$-value with the observed log Bayes factor.

\begin{figure}[!h]
  \centering%
{ \includegraphics[width=9.4cm]{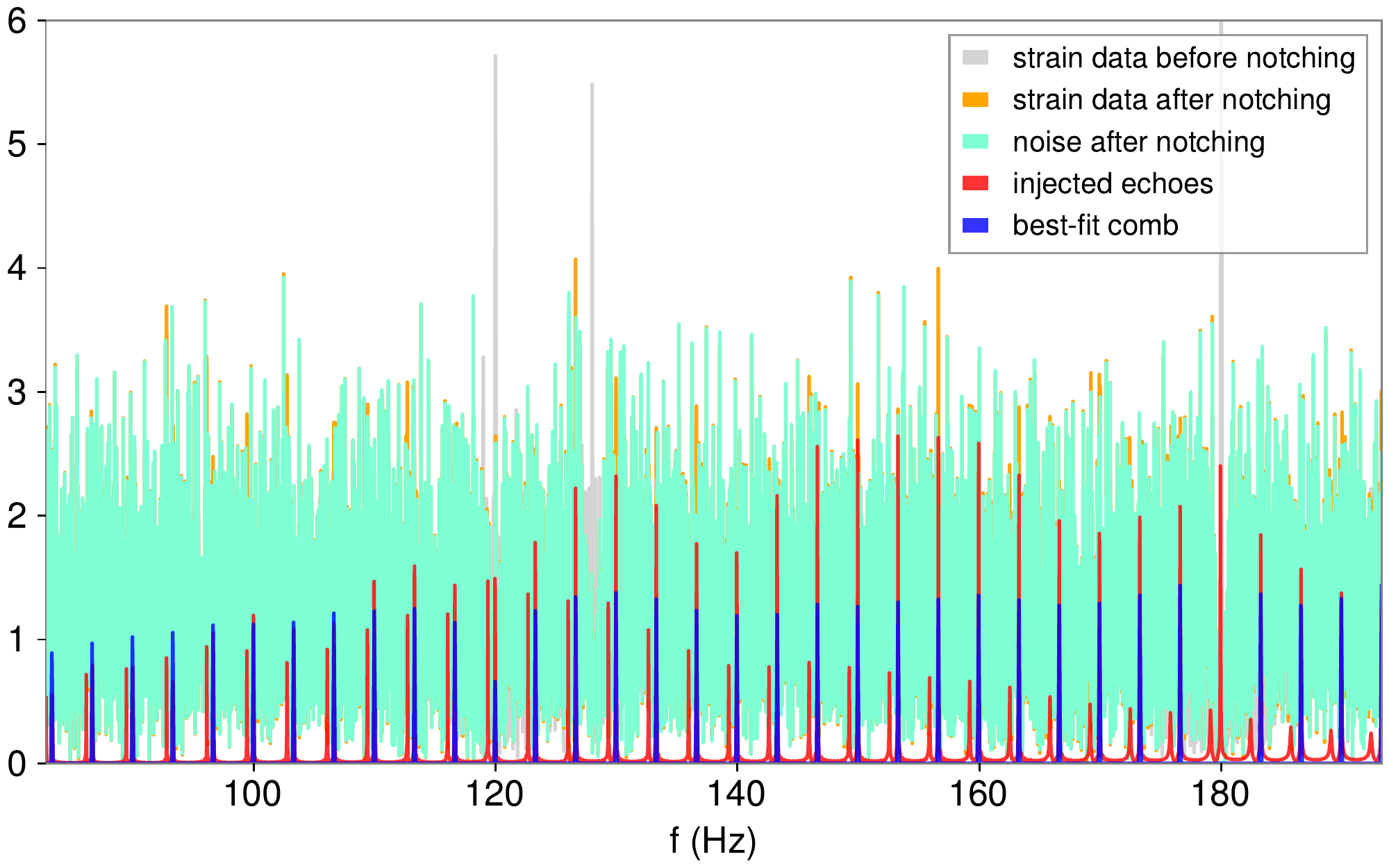}}\;
{ \includegraphics[width=6cm]{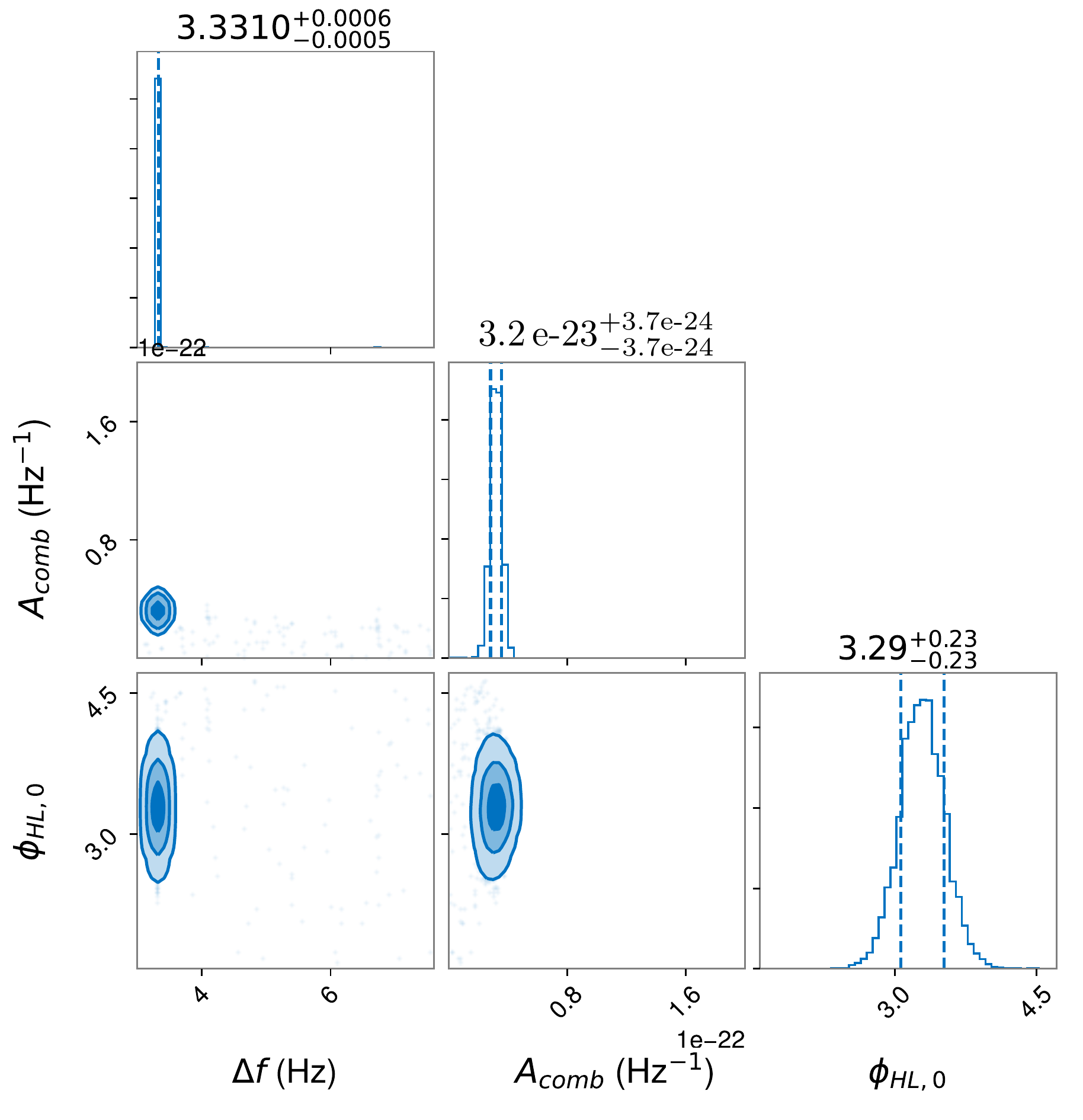}}\\
{ \includegraphics[width=7.7cm]{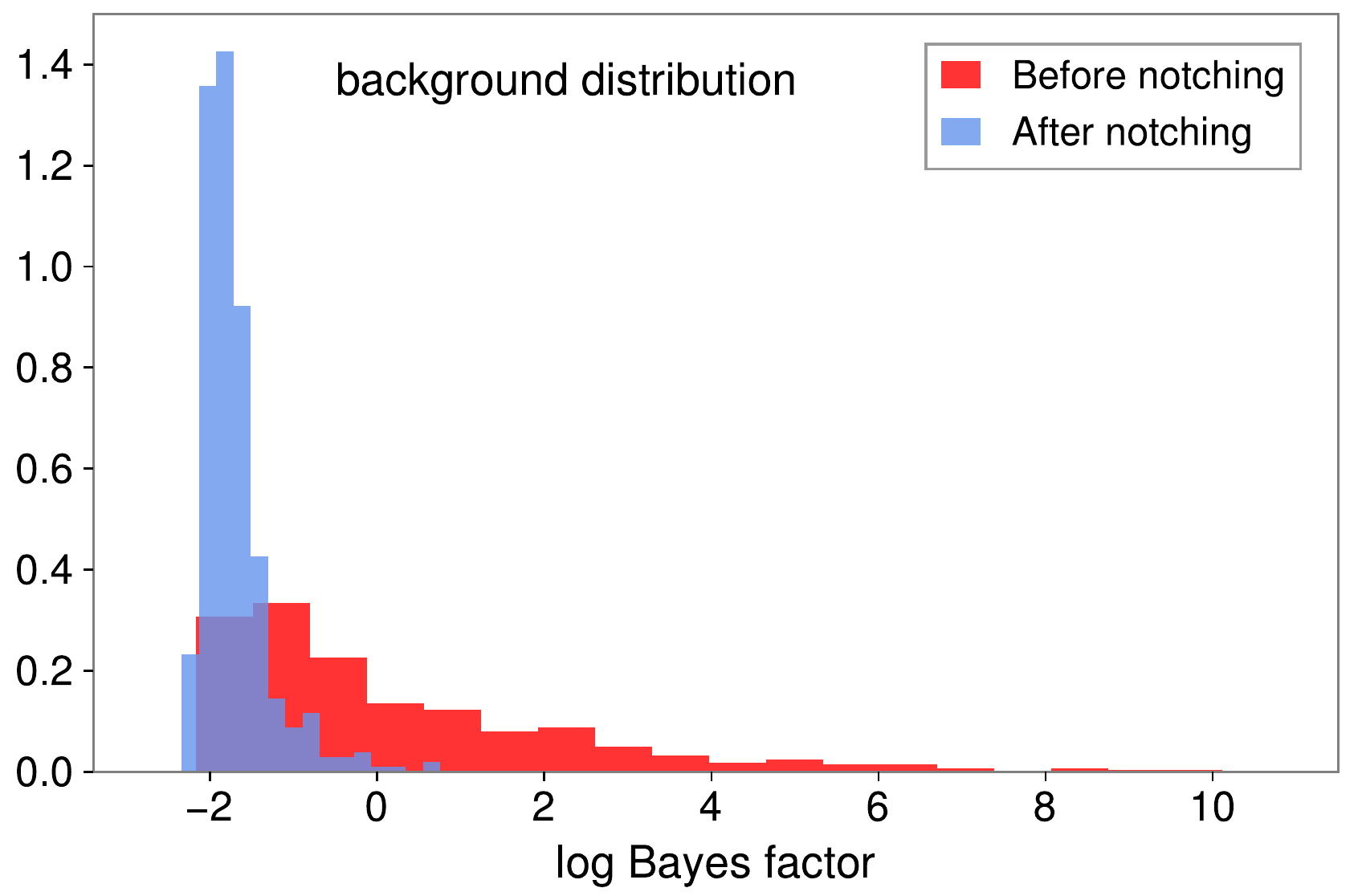}}\;
{ \includegraphics[width=7.7cm]{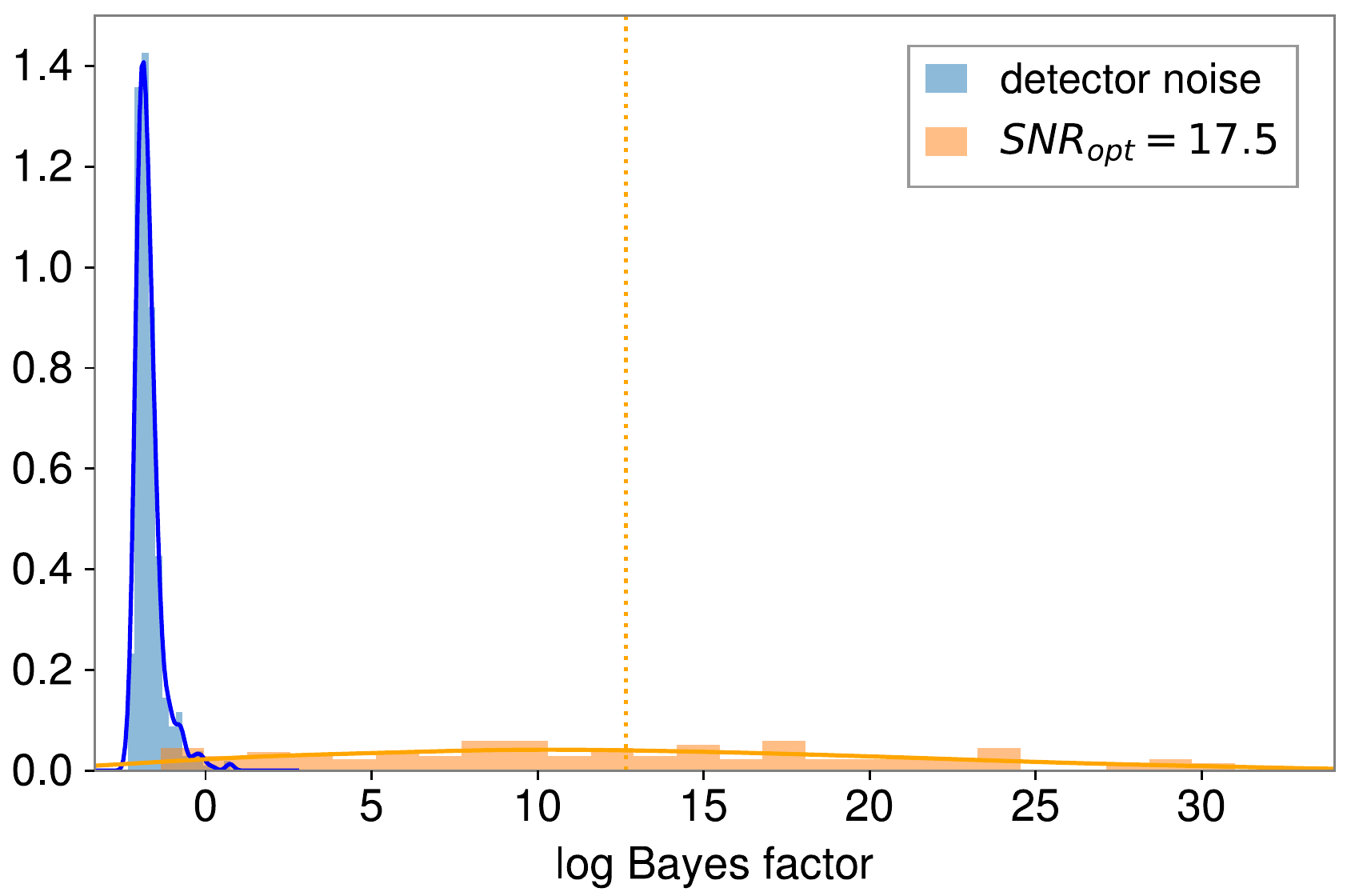}}
\caption{\label{fig:bestfit_LIGO_inject} The search results for an injected echo signal from model A in detector noise around GW150914, with $T=66.3\,$s and the average network optimal $\textrm{SNR}\approx 17.5$. Top left:  the absolute value of the normalized data in the frequency domain with $\ln\mathcal{B}\approx 2.5$. The light gray and orange (cyan) lines denote the combined strain (noise) data $\tilde{d}_{HL}$  in Eq.~(\ref{eq:rhomf2res})  before and after notching, respectively. The blue line denotes the normalized best-fit comb $\tilde{h}'$ in Eq.~(\ref{eq:rhomf2res}) with frequency bins influenced by notching removed. Top right: the corner plot for the sampled posterior distributions for half of the search parameters, where the full prior ranges are shown. Bottom left: the log Bayes factor distributions for detector noise before and after notching. Bottom right: the log Bayes factor distributions for detector noise and the injected signal after notching.}
\end{figure}

Let us start from the first confirmed event GW150914. Since the frequency band of interest is below 300\,Hz, narrow spectral artifacts of concern are the power mains at 60\,Hz harmonics and Output Mode Cleaner (OMC) length dither. 
We notch out these lines when the normalized peak values are above the threshold given in Table~\ref{tab:eventparameter}.
To further validate the search algorithm on real data analysis, we first perform the search on Advanced LIGO data with an echo signal injected into detector noise. For comparison with the search in the presence of stationary Gaussian noise, we use model A here for the injected signal with the average SNR comparable.
Figure~\ref{fig:bestfit_LIGO_inject} displays the search results. In the top-left panel, the light gray and orange lines denote normalized strain data for a sample before and after notching. As we can see, the large spectral lines at 120, 128 and 180\,Hz have been properly removed to prevent their large contribution to the likelihood. 
The bottom-left panel shows the log Bayes factor distributions for detector noise before and after notching. The distribution before notching has a long tail of noise outliers, with the inferred comb spacing at 4, $20/3$ and $15/2\,$Hz. After notching out the large spectral lines, in particular the power mains, noise outliers are properly removed, and the background distribution becomes quite similar to that in Fig.~\ref{fig:logB_comb} for stationary Gaussian noise.

The injected echo signal in the top-left panel in Fig.~\ref{fig:bestfit_LIGO_inject} has been tuned such that some resonances locate exactly at the notching frequencies. Their contributions are then not included after notching, corresponding to the missing or reduced comb teeth at 120 and 180Hz (blue line). In comparison to search results in the presence of stationary Gaussian noise, the inferred comb height is slightly larger, and this compensates the missing contribution of the notched resonances. Fortunately, the number of such resonances is small, and the injected echo can still be correctly recovered. 
From the corner plot on the right, we see that the comb spacing $\Delta f$ is determined equally well as in the case of Gaussian noise, and the error is only of the order of $0.01\%$ of the prior range. The average comb amplitude is well within the search range, demonstrating the proper choice of the prior. As for detector response, the new search parameter $ \phi_{HL,0}$ is determined to be about $10\%$ of the prior range by taking $\Delta t_{HL,0}$ at exactly the injected value.\footnote{Since the error for $\Delta t_{HL,0}$ is mostly within 1\,ms, we also check the cases when  $\Delta t_{HL,0}$ deviates from the injected value by 1\,ms. This leads to a shift of the best-fit $\phi_{HL,0}$ as shown in Fig.~\ref{fig:GW150914_inj_bestfit_dist}, while the log Bayes factor distribution as well as  the signal detection probability do not change much.}  
In comparison to Fig.~\ref{fig:logB_echo} for stationary Gaussian noise, the log Bayes factor distributions change to some extent, but not significantly. The detection probability (for a false alarm probability equal to 10\%) reduces slightly, being 95\% and  84\% for $T=66.3$ and $48.6\,$s, respectively.

\begin{figure}[!h]
  \centering%
{ \includegraphics[width=15cm]{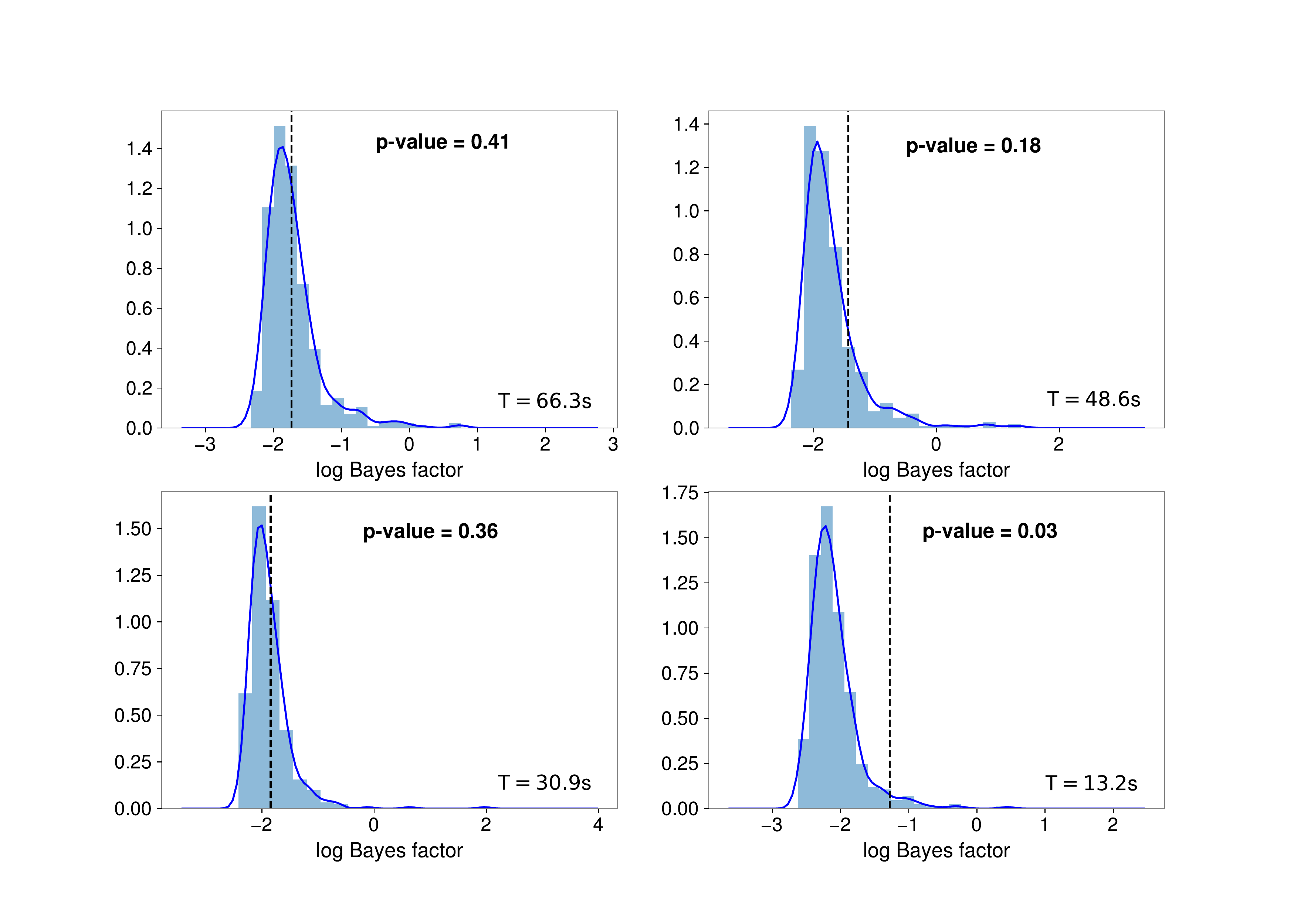}}
\caption{\label{fig:logB_GW150914} The background distribution (blue) and the observed value (dashed black line) of the log Bayes factor for GW150914. The distributions are produced by 500 samples. The four panels consider four benchmark values of the time duration $T$. The $p$-values are evaluated with the smoothened background distributions (solid blue line).}
\end{figure}

After validating the methodology and implementation of the algorithm with LIGO data, we are now ready to perform the proposed search for the confirmed gravitational wave events. 
Figure~\ref{fig:logB_GW150914} summarizes our search results for GW150914. For each time duration $T$, the blue histogram shows the background distribution of log Bayes factor with 500 samples. 
The black dashed line denotes $\ln\mathcal{B}_\textrm{obs}$, the detection statistic obtained by analyzing the data right after merger. With the smoothened background distributions, we can calculate the $p$-value with
\begin{eqnarray}
p=1-\int_{-\infty}^{\ln\mathcal{B}_\textrm{obs}}\mathcal{P}(x)dx\,,
\end{eqnarray}
 which quantifies the probability of getting $\ln\mathcal{B}$ from noise larger than that for the observed data.  The smallest $p$-value is 3\% for $T=13.2\,$s. Overall, the results are consistent with background distributions, and there is no clear evidence of a comblike structure in the strain data amplitude.

\begin{figure}[!h]
  \centering%
{ \includegraphics[width=15cm]{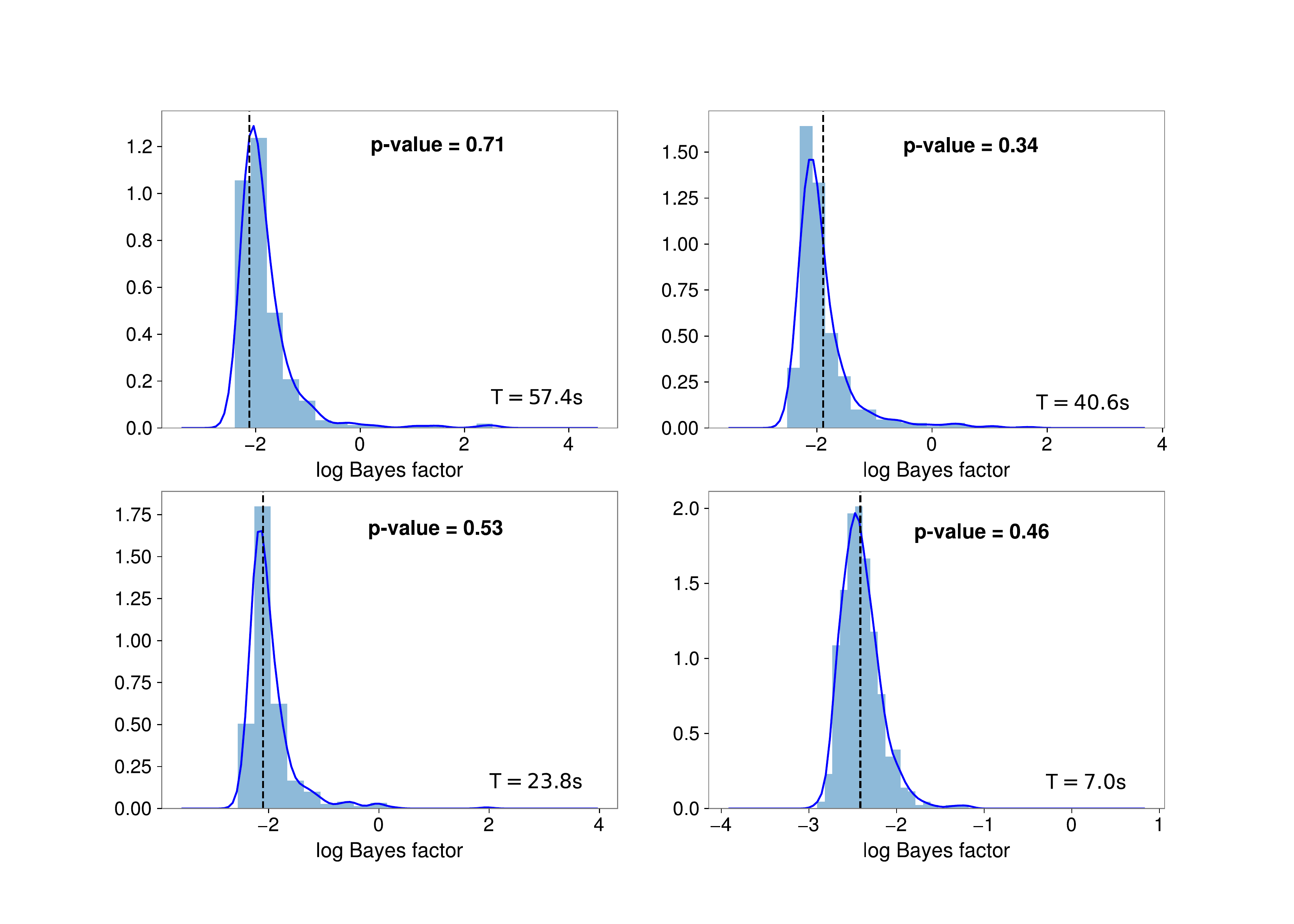}}
\caption{\label{fig:logB_GW151012} The background distribution (blue) and the observed value (dashed black line) of the log Bayes factor for GW151012. The same notation as Fig.~\ref{fig:logB_GW150914}.}
\end{figure}

We perform the search for GW151012 following the same procedure. 
As the frequency band of interest extends up to around 500\,Hz for a lighter final object, more spectral lines have to be taken into account for a proper background estimation. In addition to those considered for GW150914, we further notch out the calibration lines and the violin modes around 300\,Hz. Because of the enlarging priors for the frequency band, we set the notching threshold a bit lower and the minimum width of frequency band a bit larger as shown in Table~\ref{tab:eventparameter}.
Figure~\ref{fig:logB_GW151012} shows the search results. The overall shapes of the background distributions are similar to that for GW150914. 
The search results are consistent with background distributions, with the smallest $p$-value being 34\% for $T=40.6\,$s.

Therefore, given the priors defined in Table~\ref{tab:parameter}, we find  no clear evidence of a comblike structure as reported in the previous study~\cite{Holdom:2019bdv} for the first two events in the GWTC-1 catalog. As a simple check, we rerun the analysis on data after merger by considering an incoherent combination of two detectors to resemble more the analysis in Ref.~\cite{Holdom:2019bdv}. We also narrow down the prior ranges to be around the inferred comb parameters from the reported signals. No clear evidence shows up, indicating that the difference is unlikely to be related to the enlarging parameter space volume.  Another possible reason could be different treatments of the large instrumental lines in Advanced LIGO data, since we both see outliers with spacing at 4\,Hz before manipulation of the lines. Instead of notching, Ref.~\cite{Holdom:2019bdv} replaced large lines by the average value of noise to maintain their contributions in the test statistics.  A more close comparison of two methods is left for future work. 

\section{Summary}
\label{sec:summary}

We develop a Bayesian algorithm to search for the narrow resonance structure of the echo amplitude in the frequency domain, being complementary to the searches that focus on a small number of early time pulses. The essential inputs for the algorithm are the phase-marginalized likelihood in Eqs.~(\ref{eq:rhomf2res}) and (\ref{eq:logL2}) and a uniform comb consisting of triangular teeth, as described in Table~\ref{tab:combparameter} by a few parameters. The detector response is properly taken into account in the phase-marginalized likelihood, and a minimal set of search parameters is chosen in Table~\ref{tab:parameter} to optimize the search sensitivity. In the absence of phase information, the method is able to detect a large variety of echo waveforms in a rather model-independent way. An echo signal not much below the average noise can be recovered correctly, and the inferred comb parameters measure the essential properties of the resonance structure. The time delay $t_d$ is very well determined due to the contribution from a large number of resonances. The error and prior range ratio can be smaller than $0.1\%$ for a mild signal as shown in Figs.~\ref{fig:bestfit1} and \ref{fig:bestfit1G}.

We perform the proposed search on the strain data of the first observing run of Advanced LIGO, and the search algorithm is validated with signal injections in real detector noise. By notching out a small number of the prominent instrumental lines as reported by the collaboration, we find the background distribution of the log Bayes factor  for the comb versus noise models  quite similar to that for stationary Gaussian noise, and a large number of noise outliers are successfully removed as shown in Fig.~\ref{fig:bestfit_LIGO_inject}. 
For the injected signal, with a proper choice of the duration $T$ of the strain time series, the inferred comb parameters are determined equally well as in the case of Gaussian noise. We analyze the data right after merger for GW150914 and GW151012, the $p$-values in Figs.~\ref{fig:logB_GW150914} and \ref{fig:logB_GW151012} are consistent with the background distributions, and we find no clear evidence of a comblike structure in the strain data amplitude.

The algorithm can be further optimized as follows. First, the choice of tooth width may not be optimal. The current value at smaller $T$ is too large and the comb may include too much noise. Second, a more careful scan over the time duration $T$ could be useful. As $T$ is not implemented as a search parameter for Bayesian inference, it is impossible to reach a high precision or to impose constraints together with other parameters. Lastly, more features of the resonance structure can be implemented in the search. This includes a better choice of the prior ranges for the search parameters and the possibility of using a more complicated comb. For instance, a comb with two separated frequency bands may significantly enhance the sensitivity if the resonance amplitude is suppressed in a wider range of frequency around $f_H$ or the two components of the spectrum are more comparable in size.

With further optimization of the algorithm, we will continue searching for the resonance structure of echoes for more confirmed gravitational wave events. The second and third observing runs of Advanced LIGO have fewer instrumental lines~\cite{Covas:2018oik,Davis:2021ecd}. Also, with Virgo and KAGRA joining the global network,  the cross-detector correlation can play a more important role.  
Our algorithm can be easily generalized to perform a combined search for multiple events. If the signal is too weak to be detected for an individual event, a combined search with correlated prior ranges may still have chances.   
Finally, the resonance structure of echoes can also be searched for via SGWB~\cite{Du:2018cmp}, and it is interesting to combine the analyses of individual signals and background.

%%%%%%%%%%%%%%%%%%%%%%%%%%%%%%%%%%%%%%%%%%%%%%%%%%%%%%%%%%%%%%%%%%%%%%

\vspace{0.1cm}
\section*{Acknowledgements} 
\vspace{-0.1cm}
We  thank Niayesh Afshordi and Bob Holdom for insightful discussions of using Rice distribution to optimize the original comb method before. 
We are grateful to Bob Holdom for pointing out the connection between the power-spectral density likelihood in BILBY documentation and the Rice distribution, and for useful comments on the search method and final results. We thank Imene Belahcene for valuable discussions on the usage of Advanced LIGO data and for carefully reading the manuscript. We thank Sai Wang for the nice introduction to the BILBY library. We thank Vitor Cardoso for useful comments on the manuscript. 
We are grateful to the anonymous referee for the useful suggestions and comments.
J.~Ren is supported by the Institute of High Energy Physics, Chinese Academy of Sciences, under Contract No. 
Y9291220K2.

This research has made use of data, software and/or web tools obtained from the Gravitational Wave Open Science Center (https://www.gw-openscience.org/), a service of LIGO Laboratory, the LIGO Scientific Collaboration and the Virgo Collaboration. LIGO Laboratory and Advanced LIGO are funded by the United States National Science Foundation (NSF) as well as the Science and Technology Facilities Council (STFC) of the United Kingdom, the Max-Planck-Society (MPS), and the State of Niedersachsen/Germany for support of the construction of Advanced LIGO and construction and operation of the GEO600 detector. Additional support for Advanced LIGO was provided by the Australian Research Council. Virgo is funded, through the European Gravitational Observatory (EGO), by the French Centre National de Recherche Scientifique (CNRS), the Italian Istituto Nazionale di Fisica Nucleare (INFN) and the Dutch Nikhef, with contributions by institutions from Belgium, Germany, Greece, Hungary, Ireland, Japan, Monaco, Poland, Portugal, Spain.

%%%%%%%%%%%%%%%%%%%%%%%%%%%%%%%%%%%%%%%%%%%%%%%%%%%%%%%%%%%%%%%%%%%%%%%%%%%%%%%%%%%%%%%%%%%%%%%%%%%%%%%%%%%%%%%%%

\appendix

\section{More details for analysis of Advanced LIGO data}

\begin{table}[h]
\begin{center}
\begin{tabular}{l||l|l}
\hline\hline
&
\\[-3mm]
Parameters & GW150914 &  GW151012
\\
&
\\[-3.5mm]
\hline
$\Delta f$ & uniform in $[3.0, 7.6]\,$Hz & uniform in $[3.5, 14.3]\,$Hz
\\
$f_0$ & uniform in $[0,1]$ & uniform in $[0,1]$
\\
%\hline
$A_\textrm{comb}$ & uniform in $[10^{-25}, 2\times 10^{-22}]\,\textrm{Hz}^{-1}$ & uniform in $[10^{-25}, 2\times 10^{-22}]\,\textrm{Hz}^{-1}$
\\
%\hline
$f_\textrm{min}$ & uniform in $[50, 154]\,$Hz  & uniform in $[50, 230]\,$Hz 
\\
%\hline
$f_\textrm{max}$ & uniform in $[189, 275]\,$Hz & uniform in $[289, 433]\,$Hz
\\
%\hline
$\phi_{HL,0}$ & uniform in $[\pi/2,3\pi/2]$  & uniform in $[\pi/2,3\pi/2]$
\\
$f_w$ & fixed at $11/T$ & fixed at $11/T$
\\
%\hline
$A_{HL}$ & fixed at $1$ & fixed at $1$
\\
%\hline
$\Delta t_{HL}$ & fixed at $6.9\times 10^{-3}\,$s & fixed at $-0.6\times 10^{-3}\,$s
\\
$T$ & scan over $ [13.2, 30.9, 48.6, 66.3]\,$s & scan over $[7.0, 23.8, 40.6, 57.4]\,$s 
\\
Constraints & $f_\textrm{max}-f_\textrm{min}>10\Delta f$ & $f_\textrm{max}-f_\textrm{min}>15\Delta f$ \\
%\hline
\hline
Line origin  & power mains, OMC length dither & power mains, OMC length dither,  \\
 & & calibration lines, violin modes \\
Threshold & $[5, 5, 6, 6]$ for increasing $T$ &  $[5, 5, 5.5, 5.5]$  for increasing $T$ \\
\hline\hline

\end{tabular}
\caption{The explicit parameter settings for the two events in the real data search. The inferred properties of the final object from analysis of the inspiral-merger-ringdown signal  are taken from \cite{GWTC1, Marcoccia:2020rag}. The last two lines are for notching of spectral lines. }
\label{tab:eventparameter}
\end{center}
\end{table}

In this appendix, we explain in more detail the special treatments for Advanced LIGO data in our Bayesian search. The explicit parameter choices for the two events are summarized in Table~\ref{tab:eventparameter}. For a pair of strain data samples from H1 and L1 with time duration $T$, first, we whiten a longer segment of strain data of duration $3T$, where the sample of interest is right in the middle. The PSD used for whitening is given by matplotlib.mlab.psd with NFFT = int(0.5$f_s$), noverlap=int(NFFT/2), window=np.blackman(NFFT). A short segment of duration 0.5\,s is chosen to suppress the potential contributions from narrow signal spikes. 
Then, we create zero-pole-gain (ZPK) notch filters~\cite{gwpy} for narrow instrumental lines with the normalized peak value above some threshold,  and apply the concatenated filters to the strain data of duration $3T$. The origins of instrumental lines and the thresholds are given in Table~\ref{tab:eventparameter}.  The power mains at 60\,Hz harmonics and OMC length dither are considered for both events. For GW151012, many more lines have to be considered, including the calibration lines and the violin modes around 300\,Hz. Many violin modes are really close by, and this makes the notching procedure of GW151012 more tedious. As the lines grow for increasing frequency resolution, a larger threshold is taken for data of a longer duration $T$. 
Given the limited frequency resolution for $T\lesssim 100\,$s, extremely narrow combs with unknown origin are irrelevant to our search.  
After notching, we discrete Fourier transform the intermediate segment of whitened strain data of duration $T$, and then substitute into the normalized likelihood function in Eq.~(\ref{eq:logL2}) with the SNRs in Eq.~(\ref{eq:rhomf4}). 
Since notching modifies strain data around the lines in a range of the order of 0.1\,Hz, we remove frequency bins within $\pm 0.3\,$Hz of the notched lines entirely in evaluation of the likelihood.  
Finally, we do the Bayesian inference with the priors and fixed values of the nine parameters given in Table~\ref{tab:eventparameter}, as being derived from Table~\ref{tab:parameter}. The sampler settings remain the same as those for the Gaussian noise analysis.

\begin{figure}[!h]
  \centering%
{ \includegraphics[width=15cm]{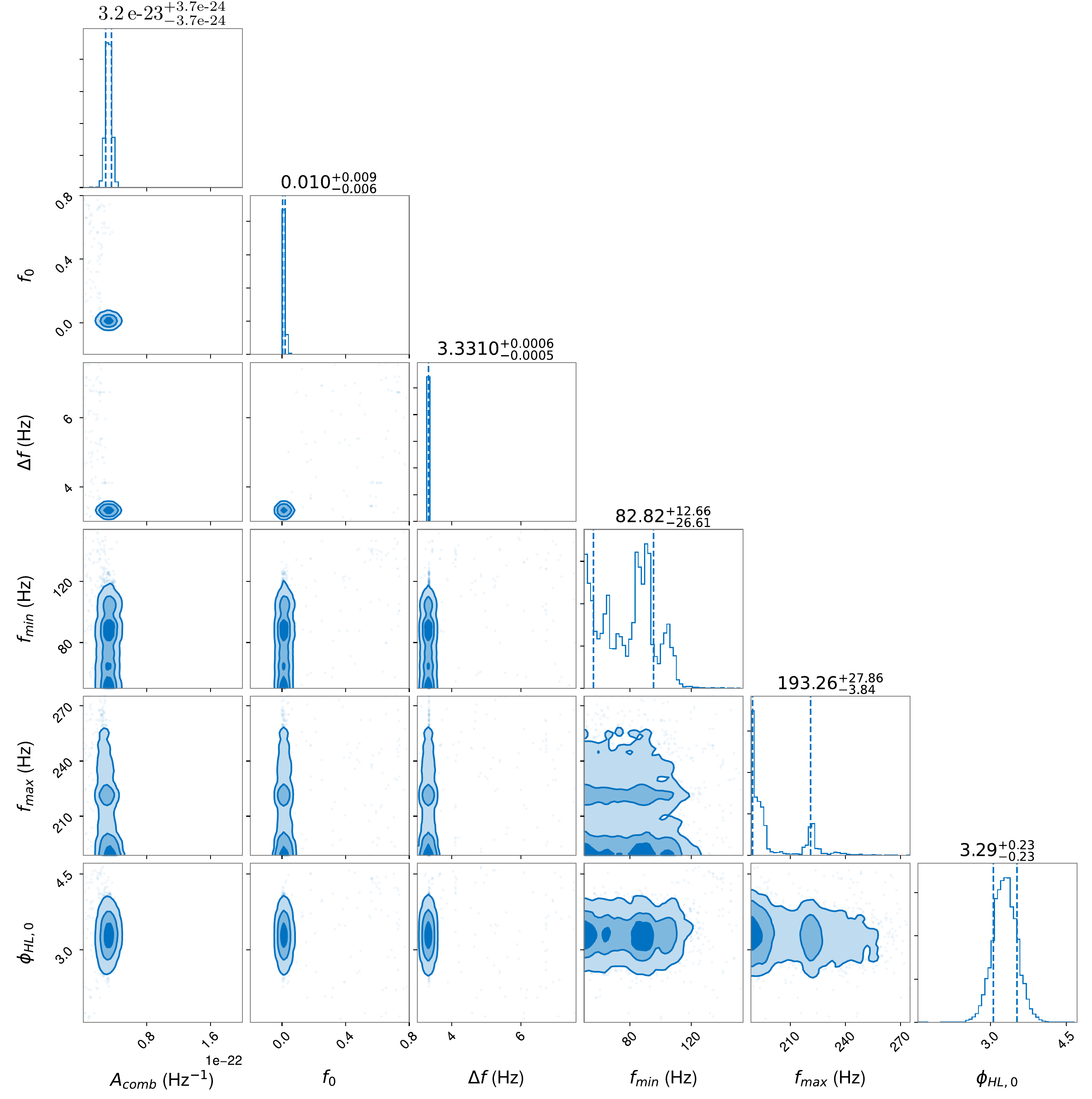}}
\caption{\label{fig:bestfit_LIGO_inject_full} The corner plot for the sampled posterior distributions of all six search parameters for the injected signal in Fig.~\ref{fig:bestfit_LIGO_inject}. The full prior ranges are shown.}
\end{figure}

For the sake of completeness, we also present more search results for the analysis of Advanced LIGO data. 
Figure~\ref{fig:bestfit_LIGO_inject_full} shows the corner plot of all six search parameters for the injected signal in Fig.~\ref{fig:bestfit_LIGO_inject}. The first three parameters $A_\textrm{comb}, f_0$ and $\Delta f$ are very well determined, while the frequency band has a large spread. The best-fit $f_\textrm{max}$ is close to the left boundary because the analytical model in Ref.~\cite{Maggio:2019zyv} has a smaller ringdown frequency due to the approximation they made. There is no strong correlation between any pair of the parameters, and this is consistent with our minimal choice of the search parameters.  
\begin{figure}[!h]
  \centering%
{ \includegraphics[width=14cm]{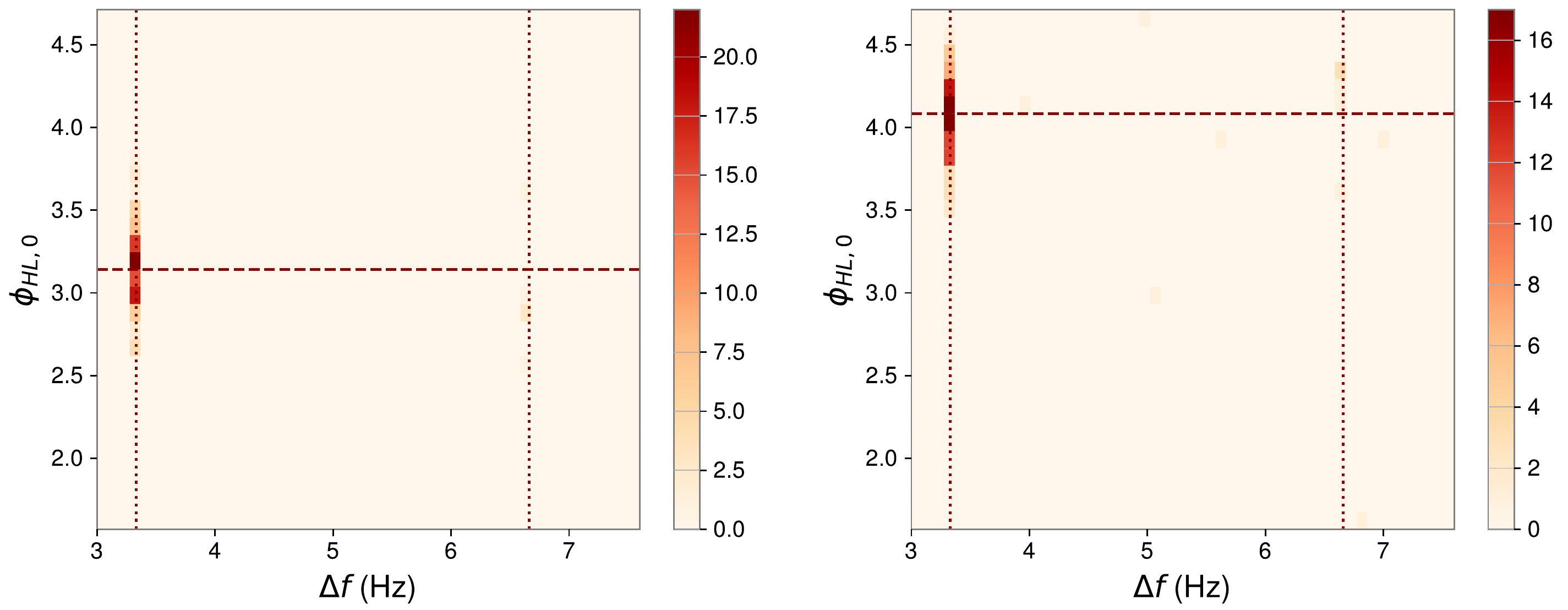}}
\caption{\label{fig:GW150914_inj_bestfit_dist} The 2D distributions of the best-fit comb spacing v.s. the constant response phase for the samples with the injected echo in Fig.~\ref{fig:bestfit_LIGO_inject}. Left: the time lag fixed the same as the injected value. Right: the time lag one millisecond larger than the injected value. The two vertical dotted lines denote the injected value and twice of the injected value of spacing. The horizontal line on the left and right panels is $\pi$ and $\pi+2\pi (150\,\textrm{Hz})(1\,\textrm{ms})$, respectively.}
\end{figure}

Figure~\ref{fig:GW150914_inj_bestfit_dist} shows the distributions of the best-fit comb spacing versus the constant response phase for samples with the same injected echo. When the time lag $\Delta t_{HL,0}$ is fixed the same as the injected value, the best-fit constant phase is around $\pi$ due to the nearly antialigned configurations of the two LIGO detectors, as shown by the horizontal dashed line in the left panel. If the time lag is one millisecond larger than the injected value, the constant phase has to be larger to keep $\phi_{HL,j}=\phi_{HL,0}-2\pi f_j \Delta t_{HL,0}$ not much changed. This shows the degeneracy between the two phases as mentioned in the main text. Since the second term does not change much within the frequency band of interest, the two cases have very similar distributions of the log Bayes factor.

%%%%%%%%%%%%%%%%%%%%%%%%%%%%%%%%%%%%%%%%%%%%%%%%%%%%%%%%%%%%%%%%%%%%%%%%%%%%%%%%%

\raggedright  
\bibliography{References_GWE}{}
\bibliographystyle{apsrev4-1}

\end{document}